\documentclass[12pt]{article}
\usepackage{amsmath,amssymb,bm,epsf,epsfig,graphicx,dsfont}

\input epsf.sty
\numberwithin{equation}{section}
\numberwithin{figure}{section}

\newcommand{\Tr}{\mathop{\rm Tr}\nolimits}

\def\bra#1{\langle #1 |}
\def\ket#1{|#1 \rangle}
\def\aver#1{\left\langle\, #1 \,\right\rangle}
\def\averfx#1{\langle\, #1 \,\rangle}

\def\Ishibashi#1{|#1 \rangle\!\rangle}
\def\no#1{:\!#1\!:}

\def\V{{\cal V}}

\def \be {\begin{equation}}
\def \ee {\end{equation}}
\def \bea {\begin{eqnarray}}
\def \eea {\end{eqnarray}}
\def \bdm {\begin{displaymath}}
\def \edm {\end{displaymath}}

\begin{document}
\vskip 2.1cm

\centerline{\Large \bf  Boundary State from Ellwood Invariants}
\vspace*{8.0ex}

\centerline{\large \rm Mat\v{e}j Kudrna$^{(a)}$\footnote{Email: {\tt matej.kudrna at email.cz}}, Carlo Maccaferri$^{(a,b)}$\footnote{Email:
{\tt maccafer at gmail.com}},
Martin Schnabl$^{(a)}$\footnote{Email: {\tt schnabl.martin at gmail.com}}
}

\vspace*{4.0ex}
\begin{center}
$^{(a)}${\it {Institute of Physics of the ASCR, v.v.i.} \\
{Na Slovance 2, 182 21 Prague 8, Czech Republic}}
\vskip .4cm

$^{(b)}${\it Dipartimento di Fisica, Universit\'a di Torino and  INFN    Sezione di Torino\\
Via Pietro Giuria 1, I-10125 Torino, Italy}
\end{center}
\vspace*{6.0ex}

\centerline{\bf Abstract}
\bigskip
Boundary states are given by appropriate linear combinations of Ishibashi states.
Starting from any open string field theory solution and assuming
Ellwood conjecture we show that every coefficient of such a linear
combination is given by an Ellwood invariant,  computed in a
slightly modified theory where it does not trivially vanish by the
on-shell condition. Unlike the previous construction of Kiermaier,
Okawa and Zwiebach, ours is linear in the string field, it is
manifestly gauge invariant and it is also suitable for solutions
known only numerically. The correct boundary state is readily
reproduced in the case of known analytic solutions and, as an
example, we compute the energy momentum tensor of the rolling
tachyon from the generalized invariants of the corresponding
solution. We also compute the energy density profile of
Siegel-gauge multiple lump solutions  and show that, as the level
increases, it correctly approaches a sum of delta functions. This
provides a gauge invariant way of computing the separations
between the lower dimensional D-branes.

 \vfill \eject

\baselineskip=15.5pt

\tableofcontents

\setcounter{footnote}{0}
\newpage
\section{Introduction}

In attempts to explore the landscape of open string field theory
\cite{Witten}\footnote{For recent reviews see e.g. \cite{TZ,FK}.}
either by analytic or numerical means, one faces the problem of a
physical identification of the solutions to the equation of
motion. They are believed to be in one-to-one correspondence with
allowed boundary states for given bulk CFT, but so far we have had
only limited tools to identify the respective boundary state
\cite{Sen:2004cq, Ellwood, KOZ}. In \cite{KOZ} a geometric
construction of the boundary state was given, in principle,
for any classical solution of open string field theory (OSFT).
However, due to the nonlinearity of
the construction, it is not known how to explicitly perform
computations for generic solutions like, for example, the
important class of Siegel-gauge level truncated solutions.
Moreover, the OSFT boundary state of \cite{KOZ} is not guaranteed
to be gauge invariant, and the BCFT boundary state is recovered
only up to BRST-exact terms (which are however absent in several
explicit examples of wedge-based analytic solutions).

In this work we present a remarkably simple method to construct
explicitly, in a gauge invariant way, the BCFT boundary state from
a given solution.
 The main advantage of the method is that, while it easily gives
the expected results for known analytic solutions, it also works
reasonably well  for solutions known only numerically. The key
ingredient of our construction is the widely believed, but as yet
unproven Ellwood's conjecture \cite{Ellwood, KOZ}, which can be
simply re-stated as
\begin{equation}
\label{Ell-conj} \aver{{\cal
V}_{cl}|\,c_0^-|B_\Psi} = -4\pi i\aver{I|{\cal
V}_{cl}(i)|\Psi-\Psi_{\mathrm{TV}}},
\end{equation}
where ${\cal V}_{cl}$ is an on-shell closed string vertex operator of ghost
number two, $\Psi$ is a solution of the OSFT, $\ket{B_\Psi}$ is
the corresponding boundary state, and finally $\Psi_{\mathrm{TV}}$
is the tachyon vacuum (in any gauge or form). Note that the left
hand side is evaluated using the closed string inner product,
while the right hand is evaluated using the open string inner
product.

This equation, however, constrains only the tiny on-shell part of
the boundary state.  For example, for spatially constant string
fields the only nontrivial component of the corresponding
boundary state which can be computed this way is just the zero
momentum massless closed string mode with vertex operator of the
form $\xi_{\mu\nu} \partial X^\mu \bar\partial X^\nu$. A possible
way around this problem has in fact been hinted at already in
\cite{KKT}, before the Ellwood's conjecture had been formulated.
The trick is essentially to assume the existence of some
spacetime direction with Dirichlet boundary condition. The vertex
operator can then be taken to have arbitrary momentum dependence
in the directions we are interested in. To put the whole operator
on-shell, we adjust the momentum of the closed string vertex
operator in the extra direction with Dirichlet boundary condition.
Due to the Dirichlet condition the invariant will not vanish
trivially. In this way, we can unambiguously extract the overlap between
the boundary state and closed string matter primaries. The goal of
this paper is to make this idea more precise and to illustrate it
on the  examples of the analytic rolling tachyon and  numerical
lump solutions describing lower dimensional D-branes.

In the matter CFT, the knowledge of the inner product of the
boundary state with any primary  state is sufficient to determine
the complete boundary state with the help of the Virasoro gluing
conditions $(L_n - \bar L_{-n}) \ket{B} =0$. These conditions are
solved in full generality by the conformal Ishibashi states
\cite{Ishibashi}, while the Ellwood invariants determine their exact
linear combination. In principle, OSFT thus solves the outstanding
unsolved problem of boundary conformal field theory: determine the
set of all boundary conditions consistent with conformal symmetry
in a given CFT. In the string theory language this is the problem
of classification of all allowed D-branes in a given background.
The coefficients of the Ishibashi states must satisfy lots of
constraints, either from modular invariance or the so called
sewing conditions. A lot of progress has been achieved in CFT
attempting to solve these constraints, but much more remains.
String field theory solutions, on the other hand, should provide
automatically a solution to all these constraints.

Let us describe the computation of the primary  components of the
boundary state in a little bit more detail. The state space of
open string field theory is given by the Hilbert space of the
boundary conformal field theory BCFT$_0$. Any such element in the
ghost number one sector can be written as \cite{Sen:1999xm, cons}
\begin{equation}\label{sol-univ}
\Psi = \sum_j
\sum_{\scriptsize\begin{array}{c}I=\{n_1,n_2,\ldots\} \\
J=\{m_1,m_2,\ldots\}\end{array}}    a_{IJ}^{j}  \,
L_{-I}^{\mathrm{matter}} \ket{\phi_j} \otimes
L_{-J}^{\mathrm{ghost}} c_1\ket{0},
\end{equation}
where the index $j$ runs over all matter primaries that are
`turned on', while the multi-indices $I$ and $J$ give its
descendants. The tachyon vacuum does not turn on any primary other
than the identity operator, while for example for the lumps an
infinite number of $e^{i k X}$  primaries (among others) is
required.

Given a solution $\Psi$ of string field theory built upon
BCFT$_0$, one can associate to it a solution $\tilde\Psi$ built
upon BCFT$_0 \otimes$BCFT$^{\mathrm{aux}}$, which only depends on
BCFT$^{\mathrm{aux}}$ through Virasoro operators and where
BCFT$^{\mathrm{aux}}$ is a (non-unitary) BCFT of central charge
$c=0$.
In simple cases (in fact in all encountered cases) this is easily
done by appropriately replacing the matter energy momentum tensor
$T^{\mathrm{matter}}$ with $T^{\mathrm{matter}}+T^{\mathrm{aux}}$.
Let us further assume that BCFT$^{\mathrm{aux}}$ contains a bulk
primary operator of dimension $(1-h,1-h)$ with nonvanishing disk
1-point function for every weight $(h,h)$ primary in the matter
part of BCFT$_0$. One universal option is to choose a BCFT of a
free boson (let us call it $Y$) with Dirichlet boundary condition
and consider generically non-normalizable operators
$e^{2i\sqrt{1-h} \,Y}$. To ensure zero central charge of
BCFT$^{\mathrm{aux}}$, one should tensor this free boson theory
with a non-unitary theory of negative central charge, for example
a $c=-1$ linear dilaton theory, and supplement the closed string
insertion with the appropriate weight (0,0) primary $w$, to soak
up the background charge.\footnote{In most cases, however, such a
construction is not necessary. It is enough to assume the
existence of a spacetime direction along which nothing happens,
and change its boundary condition to Dirichlet, if it is not
Dirichlet to start with.} Since BCFT$_0$ and BCFT$^{\mathrm{aux}}$
are completely decoupled, and the uplifted solution does not turn
on any BCFT$^{\mathrm{aux}}$ primaries other than the identity,
the boundary conditions of BCFT$^{\mathrm{aux}}$ should not be
changed by the solution. We thus expect that the boundary state
for the uplifted OSFT solution has the following factorized form
\begin{equation}
\ket{B_\Psi}^{\mathrm{CFT_0 \otimes CFT^{\mathrm{aux}}}} =
\ket{B_\Psi}^{\mathrm{CFT_0}}  \otimes
\ket{B_0}^{\mathrm{CFT^{\mathrm{aux}}}}.
\end{equation}
Assuming further that the boundary state for the solution $\Psi$ itself factorizes into matter and universal ghost parts
\begin{equation}
\ket{B_\Psi}^{\mathrm{CFT_0}}=\ket{B_\Psi}^{\mathrm{CFT_0^{matter}}}\otimes
\ket{B_{gh}},
\end{equation}
 and decomposing $\ket{B_\Psi}^{\mathrm{CFT_0^{matter}}}$ into the basis of Ishibashi states
\begin{equation}\ket{B_\Psi}^{\mathrm{CFT_0^{matter}}} = \sum_{\alpha} n_\Psi^\alpha \, \Ishibashi{V_\alpha}
\end{equation}
belonging to the (product of left and right) Verma modules of the matter
primary operators $V_\alpha$, we can determine all the coefficients
from the knowledge of the generalized Ellwood invariants
\begin{equation}
n_\Psi^\alpha =  2\pi i \aver{I|{\cal
V}^\alpha(i)|\tilde\Psi-\tilde\Psi_{\mathrm{TV}}}^{\mathrm{BCFT_0}\otimes\mathrm{BCFT^{aux}}},
\end{equation}
where $${\cal V}^\alpha = c\bar c V^\alpha\,
e^{2i\sqrt{1-h_\alpha} \,Y}\,w$$ and $V^\alpha$ form a dual basis
in the matter part of CFT$_0$, i.e.
 $$\aver{{V}^\alpha|{V}_\beta}=\delta^\alpha_\beta.$$
 This is our main result.

The paper is organized as follows. In section~\ref{s-bs}  we describe
our construction  of the boundary state in more detail. In section~\ref{s-rt}
we derive the boundary state for the rolling tachyon analytic solution,
while in the subsequent section~\ref{s-lumps} we apply our construction to single and double lump numerical solutions in open string field theory. We end up with some
conclusions and future perspectives. Appendix~\ref{a-aux} contains
an example of the auxiliary $c=0$ BCFT which we use to generalize
the Ellwood invariants. In appendix~\ref{a-cons} we derive a set
of conservation laws for the Ellwood invariant, which are very efficient
and practical, especially in numerical computations. In appendix~\ref{a-bound}
we discuss various universal properties of the boundary states in bosonic string theory.
In particular, we show that all conformal level-matched boundary states factorize into matter and universal ghost parts, and determine the precise form of the latter, including its normalization.
Finally, appendix~\ref{a-lumps} contains numerical results for several Siegel-gauge lump solutions, in addition to those discussed in section~\ref{s-lumps} of the main text.

\section{Boundary state from Ellwood invariants}
\label{s-bs}

In this section we construct a boundary state from a given OSFT
solution in two steps.  First, we  generalize Ellwood conjecture
in order to be able to use generic matter primaries in the Ellwood
invariant. Then, we show that a generic boundary state describing
conformal boundary conditions in a total matter/ghost BCFT is
necessarily matter-ghost factorized, and use the Virasoro gluing
condition of the matter sector to fix the non-primary part of the
matter boundary state. Finally, we comment on the relation between
the boundary operators turned on by the solution, and the boundary
state.

\subsection{Generalizing the Ellwood invariant}
\label{ss-gei}

Let BCFT$_0$ be the reference boundary CFT on which we define
OSFT. Let $\Psi$ be a solution  describing another BCFT$_\Psi$.
Then Ellwood conjectured  that \cite{Ellwood,KOZ}
\begin{equation}
\label{Ellconj} -4 \pi i\aver{I|\V(i,-i)|\Psi}\equiv-4\pi
i\aver{E[\V]|\Psi}=\aver{\V|\,c_0^-|B_\Psi}-\aver{\V|\,c_0^-|B_0}.
\end{equation}
Here ${\cal V}$ is a closed string vertex operator of the form
\bdm \V=c\bar c\, V^{\mathrm{matter}} \edm and \bdm
\bra{E[\V]}\equiv\bra I \V(i,-i) \edm is a corresponding state in
the open string Hilbert space. Because $\V$ is inserted at a
conical singularity (the midpoint of the identity string  field)
the quantity $\aver{I|\V(i,-i)|\Psi}$ is only meaningful  when
$\V$ is a weight zero primary. Luckily all the ghost-number two
closed string cohomology (except for the ghost dilaton) is
contained in states of this form and thus (\ref{Ellconj}) can be
used to define the on-shell part of the boundary state
$|B_\Psi\rangle$. But this is clearly not enough to completely
define the boundary state.

This is the well-known limitation of Ellwood invariants: most of
the closed string tadpoles vanish by momentum conservation when
the closed string is on-shell. This limitation lead the authors of
\cite{KOZ} to the construction of a family of  Wilson-loop-like
maps from the classical solution $\Psi$ to ghost-number-three
level-matched and BRST-invariant closed string states which are
conjectured to be BRST equivalent to the boundary state. In
particular one can probe them with off-shell closed string states.
Assuming Ellwood conjecture (or alternatively, assuming background
independence of a version of  open-closed string field theory
\cite{KOZ}), the BRST equivalence to the BCFT boundary state can
be established. The construction  is completely performed within
the open string star algebra and its intrinsic nonlinearity can
give nontrivial checks on the regularity of proposed OSFT
classical solutions \cite{Takahashi:2011wk}. But there is a
simple shortcut to get {\it precisely} the BCFT-boundary  state
described by the classical solution $\Psi$. Suppose we are dealing
with a solution which does not depend on a target space direction,
say $Y$. This means that the $Y$ dependence of the solution can be
taken to be universal, depending only on Virasoro generators of
the $Y$-BCFT. Then the solution will remain a solution if we
change the boundary conditions of the $Y$-BCFT to be Dirichlet
$Y(0)=Y(\pi)=0$. A generic closed string vertex operator of the
form $c\bar c V^{(h,h)}$, where $V^{(h,h)}$ is a bulk $(h,h)$
matter primary not depending on the $Y$ direction, can be formally
put on-shell by going to a complexified mass shell. This can be
done by multiplying $c\bar cV^{(h,h)}$ with $e^{2i\sqrt{1-h}\,Y}$
whose conformal weight is $(1-h,1-h)$. For $h>1$ (typical case)
this is a negative weight primary which is in general not
normalizable due to the divergent zero mode integration in the
world-sheet path integral. But this is not a problem with
Dirichlet boundary conditions, as the zero mode path integral will
be localized at $y=0$. Moreover, since disk one point functions of
bulk exponential operators are nonvanishing with Dirichlet
boundary conditions, the corresponding modified tadpoles will also
be generically nonzero.

\subsubsection{Lifting solutions}

The above example suggests that given a solution $\Psi$  we can
consider adding an auxiliary BCFT$^{\mathrm{aux}}$ sector of total
$c=0$  to the basis states of the original BCFT$_0$ and search for
a minimal extension or a lift of the solution, so that it becomes
a solution in the lifted OSFT defined on $$\mathrm{BCFT}_0'\equiv
\mathrm{BCFT}_0\otimes\mathrm{BCFT}_0^{\mathrm{aux}}$$ with a
lifted BRST charge
\begin{equation}
Q\to\tilde
Q\equiv\sum_n\,:c_{-n}\left(L^{\mathrm{matter}}_n+L^{\mathrm{aux}}_n+\frac12L_n^{\mathrm{ghost}}\right):.\label{lift-Q}
\end{equation}

If $\Psi$ describes new boundary conditions BCFT$_\Psi$, we search
for a solution $\tilde\Psi$ to the lifted equation
\be \tilde
Q\tilde \Psi+\tilde\Psi*\tilde\Psi=0,
\ee
such that it
describes the boundary conditions BCFT$_\Psi \otimes$BCFT$_0^{\rm
aux}$, $i.e.$ it doesn't change  the boundary conditions in the
auxiliary BCFT. We expect that this  requirement  can be achieved by imposing that
$\tilde\Psi$ depends on auxiliary degrees of freedom only through
Virasoro operators. This expectation is supported by the many
analytic
and numerical examples we have studied, although we don't have an explicit proof of this.

Our ansatz for the lifted solution is thus
\be
\tilde\Psi=\sum_{M}\Psi_M \otimes L^{\rm aux}_{-M}\ket0^{\rm
aux},\label{gen-lift}
\ee
where $M$ is a multi-index of the form
\begin{equation}
M=\{m_k,...,m_1\},\quad m_k\geq m_{k-1}\geq...\geq m_1\geq 2,
\end{equation}
and $L_{-M}$ stand
for the corresponding product of negatively moded Virasoros
\begin{equation}
{L}_{- M}\equiv L_{-m_k}...L_{-m_1}.
\end{equation}
The $\Psi_M$'s  are  ghost number
one states in the original BCFT$_0$. The above expression defines the
state $\tilde\Psi$ in terms of level expansion with respect to $L_0^{\rm aux}$. The
$c=0$ nature of the auxiliary CFT, together with the conservation
laws for the star product and the form of the lifted BRST charge
(\ref{lift-Q}), implies that the equation of motion in the tensor
theory reduces to the equation of motion in the original theory.
Concretely, whenever we have a lifted solution $\tilde\Psi$, we
can recover the original solution $\Psi$ by just looking at the
auxiliary-level-zero part of $\tilde\Psi$, that is the part of
$\tilde\Psi$ which is proportional to the auxiliary SL(2,R)
vacuum. Denoting the $L_0^{\mathrm{aux}}=0$ part of
$\tilde\Psi$ as $\Psi$
\be
\tilde\Psi=\Psi\otimes\ket0^{\rm aux}+ (L_0^{\mathrm{aux}} \ge 2 \; \mathrm{terms}),
\ee
 one can easily show that
\be \tilde Q\tilde \Psi+\tilde\Psi*\tilde\Psi=0, \ee implies \be
Q\Psi+\Psi*\Psi=0.
\ee
This is so, since the product $L_{-M}^{\mathrm{aux}} \ket{0}^{\mathrm{aux}} * L_{-N}^{\mathrm{aux}} \ket{0}^{\mathrm{aux}}$ does not contain the vacuum $\ket{0}^{\mathrm{aux}}$ unless both $M$ and $N$ are empty sets of indices.

It is tempting to think that to find such a lifted solution it
should be enough to simply change all $L^{\mathrm{matter}}$ into
$L^{\mathrm{matter}}+L^{\mathrm{aux}}$ inside the level expansion (\ref{sol-univ})
of the original solution $\Psi$. However, closer inspection
reveals that the equation of motion of string field theory is
satisfied, in general, due to cancellations between
descendants of primaries arising from conformal transformation in
the three-vertex and from the BRST charge on one hand, and
descendants appearing in the operator product expansion of primary
operators on the other hand.\footnote{Analogously, the simple
prescription $L^{\mathrm{matter}} \to
L^{\mathrm{matter}}+L^{\mathrm{aux}}$ would fail if the equations
of motion were satisfied only up to nontrivial null-vectors in the
Verma module of the identity.}

To appreciate the problem, focus on  a marginal
deformation generated by $\partial X$ to second order in the
deformation parameter $\lambda$ in Siegel gauge
\be
\Psi_\lambda=\lambda \,c\partial X(0)\ket0 - \lambda^2\frac{b_0}{L_0}\left(c \partial X \ket{0} * c \partial X \ket{0}\right)+O(\lambda^3).
\ee
To lift the
solution to BCFT$_0'$ we demand for simplicity that $\partial X$
lifts to itself (we expect that more complicated lifts do not lead
to factorized boundary state). Because of the Siegel gauge condition, the lifted solution is then uniquely specified by the first order in $\lambda$ term,
\be
\tilde\Psi_\lambda=\lambda \,c\partial X(0)\ket0 - \lambda^2\frac{b_0}{L_0+L_0^{\rm aux}}\left(c \partial X \ket{0} * c \partial X \ket{0}\right)+O(\lambda^3).\label{lift-marg}
\ee
To evaluate the
star product $c \partial X \ket{0} * c \partial X \ket{0} = {\widehat U}_3 \, \widetilde{c \partial X} (\frac{\pi}{4}) \widetilde{c \partial X} (-\frac{\pi}{4}) \ket{0}$ (for the notation see \cite{Schnabl}) one
needs the OPE of $\partial X$ with itself. To lowest order in the
level expansion one finds a coefficient times the identity operator
and in the next-to-leading order the world-sheet energy momentum
tensor $T^X$ in the free boson BCFT$^X$ with $c=1$.
In the total matter BCFT,  $T^X$ decomposes as
\be
T^X = \frac{25}{26} \left(T^X - \frac{1}{25} T'\right) +
\frac{1}{26}\left(T^X + T'\right),
\ee
where $T'$ denotes the energy momentum tensor of the rest of
matter CFT with $c=25$. The first term is a conformal primary, the
second is a descendant of the identity operator. So if we were
lifting all the descendants via $T^{\mathrm{matter}} \to
T^{\mathrm{matter}} + T^{\mathrm{aux}}$, we would have to change also the
primary $\left(T^X - \frac{1}{25} T'\right) \to \left(T^X -
\frac{1}{25} T'\right) - \frac{1}{25} T^{\mathrm{aux}}$, to keep
the OPE of $\partial X$ with itself preserved. Had we just blindly
applied  $L^{\mathrm{matter}} \to  L^{\mathrm{matter}} +
L^{\mathrm{aux}}$ to the level expansion (\ref{sol-univ}) of the original solution,
 we would have violated the equation of motion in
the lifted theory. Notice however that if we level expand the solution in the   BCFT$^{(X)}$$\otimes$BCFT$'$ basis, the geometrical lifting (\ref{lift-marg}) is equivalent to
 systematically changing all $L'_N$'s with $(L'+L^{\rm aux})_N$.

The above compatibility condition with the star product
is not so easy to solve in the most generic situation
but, at present, we have identified two (overlapping) families of
string fields which can be straightforwardly lifted.

The first class is the
algebra of wedge states with matter primary insertions, together
with insertions of the $c$-ghost with its worldsheet derivatives,
and line-integrals of the $b$-ghost\footnote{Obviously, by reparametrization,
the wedge algebra can be replaced by any other
surface state algebra.}. This family is rich enough to contain all known
analytic solutions.
A generic element of the algebra takes the form \be \Phi= \sum_i
F_i c B G_i c H_i,\label{anal-sol} \ee where the ghost number zero string fields
$F,G,H$ are star products  of elements of the wedge algebra and
matter primary insertions $\phi_i$,\be F_i=
f_1(K)\phi_1...f_{n_i}(K)\phi_{n_i}, \ee with the same generic
expression for $G$ and $H$. In this case it is not difficult to
realize that all multiplicative and differential properties are
left intact by defining the lifted string field as
\be
\tilde\Phi\equiv \Phi{\Big|}_{K\to K+K^{\rm aux}}.\label{geom-lift}
\ee
When level
expanded, the above string field has the general structure given
in (\ref{gen-lift}). Notice however that due to the non trivial
OPE between the matter insertions $\phi_i$, the level expansion of
the lifted string field cannot be obtained by just applying the
simple prescription $L^{\mathrm{matter}} \to
L^{\mathrm{matter}}+L^{\mathrm{aux}}$ to the level expansion of
the string field before the lift (\ref{sol-univ}). It is only when
no matter primary insertion enters the game (as it is for the
universal solutions) that the simple prescription
$L^{\mathrm{matter}} \to L^{\mathrm{matter}}+L^{\mathrm{aux}}$ in
the level expansion of the solution is guaranteed  to lift
solutions to solutions.

The other  family of simply liftable string fields is given
whenever the matter CFT is the tensor product of two factors
CFT$_1\otimes$CFT$_2$, with $c_{{\rm CFT}_2}\geq1$ and with only
BCFT$_1$ primaries switched on.\footnote{Aside of the induced non-diagonal primaries formed using operators from CFT$_ 1$ and Virasoro
descendants of the identity from CFT$_2$. The condition $c_{{\rm
CFT}_2}\geq1$ is there to avoid the presence of non trivial null
states on the Verma module on the identity.} At ghost number one
(relevant for string field theory solutions), this means that we
consider a subspace of states of the form
\be
\Phi=\sum_{i}\sum_{N,M,P}\Phi^i_{NMP}
L^{(1)}_{-N}L^{(2)}_{-M}L^{\mathrm{ghost}}_{-P}c\phi^{(1)}_i(0)\ket0,\label{spect-sol}
\ee
where $\phi^{(1)}_i(x)$ are primary boundary fields of
BCFT$_1$. Now, if a $\Phi$ of this form is a solution (as it is
the case, for example, for the numerical lump solutions, with the
exception of the D-instanton), the corresponding lifted solution
can be easily obtained by just replacing
\be
L^{(2)}\to
L^{(2)}+L^{\rm aux}.\label{spect-lift}
\ee
This is easily seen by noticing that, as
far as the BCFT$_2$ sector is concerned, the star product is
completely determined by the conservation laws which only depend
on the central charge, which is not changed by the lift. Again,
had some primary been switched on in BCFT$_2$, the simple lifting
procedure we just advocated would fail because the OPE between the
switched-on primaries would not be preserved by the lift, and the
star product would not commute with the lift. As a final comment
notice that, if BCFT$_2$ contains a $c=1$ free boson with Neumann
boundary conditions (as it is usually the case in known string
field theory constructions), one can equivalently change the
boundary conditions of this factor from Neumann to Dirichlet and
still have a solution to the equation of motion, without
explicitly tensoring an auxiliary BCFT. The relevant computations
for the boundary state are insensitive to whether we uplift a
spectator sector or whether we change boundary condition of a
spectator direction to Dirichlet. We will use the latter simple
shortcut in some  explicit examples later in the paper.

To summarize, we search for a lifted solution of the form
(\ref{gen-lift}). We don't have a completely general
analytic procedure to get the higher lifted components
$\Psi_M$, if nothing is assumed on the starting solution $\Psi=\Psi_{\emptyset}$. However,
in the case the solution $\Psi$ is made of surfaces with matter
primary insertions, then the simplest lift is given by (\ref{geom-lift}), and the corresponding coefficients $\Psi_M$ can be systematically computed
by usual methods, if one needs to. Alternatively, when the solution lives in a BCFT which is the tensor product of two factors, and no primaries are switched on
 in one of the two factors (the spectator sector), as in (\ref{spect-sol}), then the level expansion of the lifted solution is given by (\ref{spect-lift}). Notice that the two lifts coincide for
 analytic solutions of the form (\ref{anal-sol})
 with a spectator sector.

A more general explicit constructive procedure for the $\Psi_M$'s  is needed,
for example, for numerical marginal deformations along generic
directions or more fundamentally for numerical D-instanton lumps,
where there are no spectator dimensions nor a simple geometric
picture for the solution.
In such a case, however, one can construct the uplifted solution numerically. Since the original equations of motion are a  subset of the lifted equations of motion, level by level,
 we can uniquely link the numerical solutions of the lifted theory  to the corresponding solutions of the original theory
 by matching the coefficients in the $L_0^{\rm aux}=0 $ sector.
This is an explicit construction of the lift for the numerical solution which, however, we have not yet tested against explicit examples.

\subsubsection{Lifting closed string states}
Given any CFT bulk primary of the form
\begin{equation}
\V^\alpha(z,\bar z)=c\bar c V^\alpha(z,\bar z),
\end{equation}
where $V^\alpha$ is a purely matter primary  of weight
$(h_\alpha,h_\alpha)$, we can consider a formal bulk primary in
CFT$^{\mathrm{aux}}$, $w^\alpha(z,\bar z)$ of weight
$(1-h_\alpha,1-h_\alpha)$ with the property that
\begin{equation}
\langle\;
w^\alpha(0)\;\rangle^{\mathrm{BCFT}^{\mathrm{aux}}}_{\mathrm{disk}}=1,\quad\forall
\alpha.\label{tad-aux}
\end{equation}
Explicitly, as discussed in more detail in appendix~\ref{a-aux},
we can define $\mathrm{BCFT}^{\mathrm{aux}}$ to be the tensor
product of a free boson $Y$ with Dirichlet boundary conditions
($c=1$) and a linear dilaton $\varphi$ with background charge
$Q=\frac1{\sqrt3}$ with Neumann boundary conditions and
$c=1-6Q^2=-1$. In this case we can systematically take
\begin{equation}
w^\alpha=e^{2i\sqrt{1-h_\alpha}\,Y}e^{\frac{2i}{\sqrt3}\varphi},\label{w-example}
\end{equation}
which has weight $(1-h_\alpha,1-h_\alpha)$ and satisfies
(\ref{tad-aux}),  thanks to the Dirichlet conditions for $Y$ and
the saturation of the background charge on the disk. Notice that,
for $h_\alpha>1$,  $w^\alpha$ is not normalizable in the auxiliary
closed string Hilbert space, but still it has a well defined
one-point function on the disk. Other choices of
BCFT$^{\mathrm{aux}}$ are clearly possible.

\subsubsection{Generalized Ellwood invariant}
For OSFT purposes the closed string insertion
$$\tilde \V^\alpha\equiv c\bar c\,V^{\alpha}\otimes w^\alpha$$
will be a total $(0,0)$ bulk primary (in fact, a formal, not
normalizable,  element of the $\tilde Q$ closed string
cohomology). Thus, {\it assuming Ellwood conjecture}, the Ellwood
invariant will compute the difference in the tadpoles between the
two BCFT's related by the classical solution. But since the
solution $\tilde\Psi$ does not switch on any new primaries in
$\mathrm{BCFT}_0'$,
the generalized Ellwood invariant will be proportional to the disk
one-point function of $w^\alpha$
\begin{eqnarray}
&&-4\pi i\averfx{E[{\cal\tilde
V^\alpha}]|\tilde\Psi}^{\mathrm{BCFT_0'}}\nonumber\\&&= \langle
\tilde \V^\alpha|\,c_0^{-}\,|\tilde B_{\tilde\Psi}\rangle-
\langle\tilde \V^\alpha|\,c_0^{-}\,|\tilde B_0\rangle\nonumber\\
&&={\Big(}\langle c\bar cV^\alpha|\otimes\langle
w^\alpha|{\Big)}\,c_0^{-}\,{\Big(}|
B_{\Psi}\rangle\otimes\ket{B_{\rm aux}}{\Big)}-
{\Big(}\langle c\bar cV^\alpha|\otimes\langle w^\alpha|{\Big)}\,c_0^{-}\,{\Big(}| B_{0}\rangle\otimes\ket{B_{\rm aux}}{\Big)}\nonumber\\
&&={\Big(}\aver{c\bar c V^\alpha|\,c_0^{-}\,|B_\Psi}-
\aver{c\bar cV^\alpha|\,c_0^{-}\,|B_0}{\Big)}\;\times\;\langle\; w^\alpha(0)\;\rangle^{\mathrm{BCFT}^{\mathrm{aux}}}_{\mathrm{disk}}\nonumber\\
&&=\aver{c\bar cV^\alpha|\,c_0^{-}\,|B_\Psi}-\aver{c\bar
cV^\alpha|\,c_0^{-}\,|B_0}.
\end{eqnarray}
Notice that the auxiliary CFT disappeared from the RHS.
Conveniently, we can relate the BCFT$_0$-boundary state with the
Ellwood invariant of the lifted tachyon vacuum, $\tilde\Psi_{TV},$
and we can  write the `generalized' Ellwood conjecture in the
simple form
\begin{equation}
\boxed{\phantom{\Biggl(}~ \aver{c\bar
cV^\alpha|\,c_0^{-}\,|B_\Psi}=-4\pi i\averfx{E[{\cal\tilde
V^\alpha}]|\tilde\Psi-\tilde\Psi_{TV}}. ~~}
\end{equation}

String field theory solutions related by gauge transformations
should  describe the same BCFT and thus the same  boundary state
$\ket{B_\Psi}$. Although the right hand side is manifestly
invariant under the gauge transformations in the new OSFT based on
BCFT$_0'$, to show that it is invariant also under the gauge
transformation in the original OSFT based on BCFT$_0$ requires a
little thought. One has to show that the lifting from BCFT$_0$ to
BCFT$_0'$ commutes with gauge transformations. This is easily
arguable in the following way. Suppose we have a solution $\Psi$
and its lift $\tilde\Psi$ written as (\ref{gen-lift}). It is not
difficult to realize that if we change $\Psi$ by a gauge
transformation with group-element $U$
\be
\Psi'=U^{-1}(Q+\Psi)U,
\ee
we can very easily get an infinite family of lifted solutions
of the form (\ref{gen-lift}) which are gauge equivalent to
$\tilde\Psi$
\be
\tilde\Psi'=\tilde U^{-1}(\tilde
Q+\tilde\Psi)\tilde U.
\ee
It is enough to choose\footnote{This
also shows that the lift is not unique: for any given solution
$\Psi$ and its lift $\tilde\Psi$ one can always change the
higher level components in the auxiliary sector of $\tilde\Psi$ with a gauge
transformation $\tilde U$ whose $L_0^{\mathrm{aux}}=0$ component does not change
$\Psi$. The question remains if there are multiple
liftings  of the form (\ref{gen-lift}) of the same solution which
are not gauge equivalent and which might then give rise to
different observables. On physical grounds we expect that this cannot happen.}
\be
\tilde U=U\otimes\ket0^{\rm aux}+\sum_{M\neq\emptyset} U_M\otimes L^{\rm
aux}_{-M}\ket0^{\rm aux},
\ee
where the higher auxiliary components $U_M$ are $generic$ ghost number zero string fields in BCFT$_0$ (with the only obvious requirement that they must be
chosen in such a way that $\tilde U$ is invertible).
The $c=0$ nature of BCFT$^{\rm
aux}$ assures that both the lifted BRST charge $\tilde Q$ and the
star product behave in such a way that the auxiliary-level-zero
element of $\tilde\Psi'$ is nothing but $\Psi'$
\be
\tilde
U^{-1}(\tilde Q+\tilde\Psi)\tilde
U=U^{-1}(Q+\Psi)U\otimes\ket0^{\rm aux}+ (L_0^{\mathrm{aux}} \ge 2 \; \mathrm{terms}).
\ee
Therefore gauge equivalent classes of solutions lift to
gauge equivalent classes of lifted solutions.

\subsection{Ellwood invariants and Ishibashi states}

Using the Ellwood conjecture, we can compute the overlap \bdm
\aver {{B_\Psi}|c_0^-|\cal V} \edm for any liftable OSFT solution
$\Psi$ and closed string state $\ket {{\cal V}}$ of the form
\begin{equation}
\ket {{\cal V}}=V^{(h,h)}(0)c_{1}\bar c_{1}\ket{0}_{SL(2,C)},
\end{equation}
where ${ V}^{(h,h)}(z,\bar z)$ is a weight $(h,h)$ bulk primary in
the matter sector. Such a state necessarily obeys
\begin{eqnarray}
L_{n\geq 1}^{\mathrm{matter}}\ket \V&=&\bar L_{n\geq 1}^{\mathrm{matter}}\ket \V=0,\\
c_{n\geq 1}\ket \V&=&\bar c_{n\geq 1}\ket \V=0,\\
b_{n\geq 0}\ket \V&=&\bar b_{n\geq 0}\ket \V=0.
\end{eqnarray}
Clearly, the closed string states $c \bar c V^{(h,h)}$ do not span
the whole set of off-shell closed string fields. However, any
generic off-shell closed string state can be obtained by acting
with ghost oscillators and matter Virasoro generators on the
ground states given by the $\ket \V$'s. This choice of basis is
quite convenient from the OSFT point of view, since the states
$\ket \V$ directly enter in the Ellwood invariant. But it is also
very convenient from the closed string point of view: the
knowledge of the overlap $\bra\V c_0^{-} \ket {B_\Psi}$ is just
enough to define {\it all} overlaps with the boundary state.

This is because the boundary state is a ghost number three closed
string state which describes  conformal boundary conditions  in
the total matter and ghost Hilbert space. This is summarized by
\bea
b_0^-\ket{B_\Psi}&=&0,\\
(L_n^{\mathrm{tot}}-\bar L_{-n}^{\mathrm{tot}})\ket{B_\Psi}&=&0,\\
(Q_{gh}-3)\ket{B_\Psi}&=&0,
 \eea
where $Q_{gh}$ is the total ghost number operator obeying $
Q_{gh}\ket0_{SL(2,C)}=0$. We show in appendix~\ref{a-bound} that
these three (sets of) conditions by themselves already imply the
standard gluing conditions \bea
\left(b_n-\bar b_{-n}\right)\ket {B_{\Psi}}&=&0,\\
\left(c_n+\bar c_{-n}\right)\ket {B_{\Psi}}&=&0,\\
\left(L^{\mathrm{matter}}_n-\bar
L^{\mathrm{matter}}_{-n}\right)\ket {B_\Psi}&=&0,\\
(Q+\bar Q)\ket{B_\Psi}&=&0. \eea

These gluing conditions allow to trade raising operators acting on
the closed string state $\ket \V$ for lowering operators which
will vanish upon acting on $\ket \V$. Thus any overlap of a closed
string state $\ket W$ built by acting with raising operators on
$\ket \V$ will be proportional to the corresponding overlap of the
boundary state with $\ket \V$ itself, the constant of
proportionality being a number which can easily be computed using
the matter Virasoro algebra and the $b,c$ oscillator algebra. {\it
Thus, up to automatic operations, the boundary state for a
solution $\Psi$ is completely encoded in the (generalized) Ellwood
invariants.}

We can beautifully and very efficiently formulate these
observations in terms of the so-called Ishibashi states. Let
$\{V_\alpha\}$ be the collection of non-singular {\it spinless}
bulk  primaries of weight $(h_\alpha,h_\alpha)$ in the matter
CFT\footnote{In CFTs on noncompact target spaces $\alpha$ will in
general be a continuous variable, like the momentum. }
\begin{eqnarray}
(L_0-\bar L_0)\ket{V_\alpha}&=&(h_\alpha-h_\alpha)\ket{V_\alpha}=0\\
L_{n}\ket{V_\alpha}&=&\bar L_{n}\ket{V_\alpha}=0,\quad n\geq1.
\end{eqnarray}
Let's define a BPZ-dual basis of primaries $\{V^\beta\}$ such that
\begin{equation}
\bra {V^\alpha}V_\beta\rangle=\delta^{\alpha}_\beta.
\end{equation}
This is possible once  singular (null) states have been projected
out. To any spinless vertex operator $V_\alpha$ we can associate
the corresponding   conformal Ishibashi state, which (up to
normalization) is the unique state
 $\Ishibashi{V_\alpha}$
in the Virasoro Verma module of $V_\alpha$ satisfying the Virasoro
gluing conditions
\begin{equation}
(L_{n}-\bar L_{-n})\Ishibashi{V_\alpha}=0.
\end{equation}
The explicit form of the Ishibashi state $\Ishibashi{V_\alpha}$ to
any desired level can be found easily by solving the gluing
conditions in the Verma module of $V_\alpha$ level by level, and
one finds in the absence of null states
\begin{eqnarray}\label{IshibashiExpl}
\Ishibashi{V_\alpha} =\!&{\Bigg[}&\!1+\frac1{2h_\alpha}L_{-1}\bar L_{-1}\\
\!&&\!+B(h_\alpha,c)\left(2(1+2h_\alpha) L_{-2}\bar
L_{-2}-3(L_{-2}\bar L_{-1}^2+L_{-1}^2\bar L_{-2})+ \frac{8
h_\alpha+c}{4h_\alpha}L_{-1}^2\bar L_{-1}^2\right)\nonumber\\&&+
\cdots{\Bigg]}\ket{V_\alpha},
\nonumber\\
B(h_\alpha,c)&=&\frac1{2h_\alpha(8h_\alpha-5)+c(2h_\alpha+1)}.
\end{eqnarray}
Had there been a null state at some level, the coefficients in
this expression  at that level would be divergent. For example the
level 2 null state appears exactly for those values of $h$ and $c$
for which $B(h,c)$ diverges.  In such a case one should exclude
the null states from the Verma module.  Solving then the gluing
conditions with null states projected out gives analogous
expression to (\ref{IshibashiExpl}) but with finite coefficients.
A simple closed form of the solution to the gluing condition for
the general case has been found by Ishibashi~\cite{Ishibashi}
\begin{equation}
\Ishibashi{V_\alpha} = \sum_n \ket{n,\alpha} \otimes
\overline{\ket{n,\alpha}},
\end{equation}
where the sum runs over orthonormal basis of states in the
irreducible representation  of the chiral Virasoro algebra built
over the primary $V_\alpha$. In writing this, we have assumed that
the closed string primary $V_\alpha$ can be decomposed into the
product of holomorphic and antiholomorphic parts. Relaxing this
assumption \cite{Rozali}, we can rewrite it equivalently as
\begin{equation}
\Ishibashi{V_\alpha} =\sum_{IJ} M^{IJ}(h_\alpha) L_{-I}\bar L_{-J}
\ket{V_\alpha},
\end{equation}
where the indices $I,J$ label the non-degenerate descendants in
the conformal family of $V_\alpha$,  and  $M^{IJ}(h_\alpha)$ is
defined as the inverse of the real symmetric matrix
\begin{equation}
M_{IJ}(h_\alpha)=\bra{V^\alpha}L_{I}L_{-J}\ket{V_\alpha}=\bra{V^\alpha}\bar
L_{I}\bar L_{-J}\ket{V_\alpha}.
\end{equation}
The normalization has been chosen so that
\begin{equation}
\langle V^\alpha
\Ishibashi{V_\beta}=\bra{V^\alpha}V_\beta\rangle=\delta^\alpha_\beta.
\end{equation}
Any boundary state $\ket{ B_*}$ in the matter CFT can therefore be
written as
\begin{equation}
\ket {B_*}=\sum_\alpha n_*^\alpha \Ishibashi{V_\alpha}.
\end{equation}
If we want to {\it define} BCFT$_*$ through its boundary state,
the coefficients $n_*^\alpha$  must be precisely chosen in order
to satisfy Cardy conditions (open string analog of modular
invariance) and sewing conditions (factorization of bulk $n$-point
functions in open and closed string channels), see e.g.
\cite{Gaberdiel}. If the boundary state $\ket {B_*}$ is known, we
can easily get $n_*^\alpha$ from
\begin{equation}
n_*^\alpha=\bra{V^\alpha}B_*\rangle.
\end{equation}
In the OSFT approach to BCFT, instead of searching for linear
combinations of Ishibashi states obeying nontrivial consistency
conditions, we search for solutions to the equation of motion. If
OSFT is a consistent theory we  expect such classical solutions to
automatically describe consistent boundary conditions.

With the above premises, our proposal can be compactly written as
\begin{equation}
\boxed{\phantom{\Biggl(}~
\ket{B_\Psi}=\sum_{\alpha}n_\Psi^\alpha\; \Ishibashi{V_\alpha}
\otimes \ket{B_{gh}}, ~~}
\end{equation}
\begin{equation}
\boxed{\phantom{\Biggl(}~ n_\Psi^\alpha\equiv 2\pi i\;\aver{ E[
\tilde \V^\alpha]\big|\tilde\Psi-\tilde\Psi_{TV}},\label{KMS} ~~}
\end{equation}

where we used, see also appendix~\ref{a-bound},
\begin{eqnarray}
\bra0c_{-1}\bar c_{-1}c_0^-\ket {B_{gh}}&=&\aver{(c_0-\bar c_0)c\bar c(0)}^{\rm ghost}_{\rm disk}=-2, \label{Ngh}\\
{\cal V^\alpha}&=&c\bar c\, V^{\alpha}, \\
\tilde{\cal V}^\alpha&=&{\cal V^\alpha}\otimes w^\alpha,
\end{eqnarray}
and $\Psi_{TV}$ is any OSFT solution for the tachyon  vacuum whose
contribution replaces the corresponding contribution from the
BCFT$_0$ boundary state.

\subsection{Ellwood invariants and boundary primaries}
\label{ss-EIBP}

It is useful and interesting to elucidate the relation between the
primary boundary fields that are
 `switched on' in an OSFT classical solution and  the boundary state which is
associated to the solution via our construction through the
Ellwood conjecture. To this end  we consider the solution
expressed as
\begin{equation}\label{lev-sol}
\Psi-\Psi_{TV}=\sum_j  \sum_{I,J}\,a^j_{IJ}\, L^{\mathrm
{matter}}_{-I}\ket{\phi_j}\otimes  L^{\mathrm
{ghost}}_{-J}c_1\ket0,
\end{equation}
where $I$,$J$ are multi-indices of the form
\begin{equation}
N=\{n_k,...,n_1\},\quad n_k\geq n_{k-1}\geq...\geq n_1\geq 1,
\end{equation}
and ${L}^{\mathrm{matter}}_{-I},\,{L}^{\mathrm{ghost}}_{-J}$ stand
for the corresponding products of negatively moded Virasoros
\begin{equation}
{L}_{- N}\equiv L_{-n_k}...L_{-n_1}
\end{equation}
acting respectively on the matter primary $\ket{\phi_j}$ and the
unique ghost primary $c_1 \ket{0}$ at ghost number one.  The lifted solution in
BCFT$_0'$=BCFT$_0\otimes$BCFT$^\mathrm{aux}$ will be  given by
\begin{equation}
\tilde\Psi-\tilde\Psi_{TV}=\sum_j \sum_{I,J,M}\,a^j_{IJM}\,{
L}^{\mathrm{matter}}_{- I}\ket{\phi_j}\otimes
{L}^{\mathrm{ghost}}_{- J}c_1\ket0\otimes L^{\rm
aux}_{-M}\ket{0}^{\rm aux},\label{lift-sol-exp}
\end{equation}
where the lifted coefficients $a_{IJM}$ reduce to the original ones $a_{IJ}$ when the multi-index $M$ is the empty set.

In computing generalized Ellwood invariant
$\aver{E[\tilde\V^\beta]\Big|\tilde\Psi-\tilde\Psi_{TV}}$
associated to the closed string field
\begin{equation}
\tilde \V^\beta=\V^\beta\otimes w^\beta= c\bar c V^\beta\otimes
w^\beta,
\end{equation}
where $w^\beta$ is a weight $(1-h_\beta)$ primary in the  $c=0$
BCFT$^{\mathrm{aux}}$ with one-point function on the disk
normalized to unity (see appendix~\ref{a-aux}), it is useful to
consider the conservation laws of the anomalous derivations
\begin{equation}
K_{n}=L_n-(-1)^n L_{-n},
\end{equation}
in matter, ghost and auxiliary sectors separately. These
conservation laws can be found in \cite{KKT}
and an alternative simple derivation is offered in
appendix~\ref{a-cons}. For instance, in the matter sector, the law takes the form
\begin{eqnarray}
\bra{E[\tilde \V^\beta]}K^{\mathrm{matter}}_{2n+1} &=& 0 \nonumber
\\
\bra{E[\tilde \V^\beta]}K^{\mathrm{matter}}_{2n} &=&  n(-1)^n (13- 16 h_\beta)\,\bra{E[\tilde \V^\beta]}.
\end{eqnarray}
Thanks to these conservation laws one can
get rid of all the Virasoros in the solution, level by level. It
follows that\footnote{Since we are computing an Ellwood invariant,
the tachyon vacuum solution  can be traded for the simple string
field
 $$\tilde\Psi_{TV}\,\to\,\frac2\pi c_1\ket0.$$ Thus, in the Ellwood
invariant, the difference between $\Psi$ and $(\Psi-\Psi_{TV})$
only appears as a universal shift in the coefficient of the zero
momentum tachyon. Notice in particular that for the perturbative
vacuum we get $A_{PV}^{\beta\,\mathds{1}}= -\frac2\pi$ as the only
nonvanishing coefficient.}
\begin{eqnarray}
\aver{E[\tilde
\V^\beta]\Big|\tilde\Psi-\tilde\Psi_{TV}}=\sum_j\,A_\Psi^{\beta
j}\aver{E[\tilde \V^\beta]\Big|c\phi_j}^{\mathrm{BCFT}_0'},
\end{eqnarray}
where $A_\Psi^{\beta j}$ is a {\it gauge invariant} linear
combination\footnote{The gauge invariance of  the coefficients
follows from the gauge invariance of the left hand side, assuming
that no two boundary operators have the same bulk-boundary
two-point function for every bulk operator.} of the coefficients
$a^j_{IJM}$ appearing in the lifted solution (\ref{lift-sol-exp}).
Using the conservation laws of appendix B recursively we can write
them explicitly as
\begin{equation}
A^{\beta j}_\Psi=\sum_{I,J,M} K^{(h_\beta,h_j)}_{IJM}\,a^j_{IJM}.
\end{equation}
Notice that the constants $K^{(h_j,h_\beta)}$ depend only on the
weights of the matter closed string state $V^\beta$ and of the
boundary primary state $\phi_j$, and otherwise are completely
universal.
Obviously, the coefficients $A_\Psi^{\beta j}$ do not depend on
whether the solution $\Psi$ is expressed in the basis of Virasoro
generators, or matter and ghost oscillators. In the latter case,
analogous formulas can be obtained by applying the oscillator
conservation laws, which we derive in appendix~\ref{a-cons} too.
The final result is exactly the same, level by level, as the one
obtained using Virasoro generators.

Computing the Ellwood invariant we find
\begin{eqnarray}
-4\pi i\aver{E[\tilde \V^\beta]\Big|c\phi_j}^{\mathrm{BCFT}_0'}&=&
-4\pi i\left(\frac2\pi\right)^{h_j-1}\langle \tilde\V^\beta(i\infty)c\phi_j(0)\rangle_{C_1}^{\mathrm{BCFT}_0'}\nonumber\\
&=&-4\pi i\left(\frac2\pi\right)^{h_j-1}\frac1{2\pi i}|2\pi i|^{h_j}\langle \tilde\V^\beta(0)c\phi_j(1)\rangle_{\mathrm{disk}}^{\mathrm{BCFT}_0'}\nonumber\\
&=&-\pi\,4^{h_j}\langle \tilde\V^\beta(0)c\phi_j(1)\rangle_{\mathrm{disk}}^{\mathrm{BCFT}_0'}\nonumber\\
&=&\pi\, 4^{h_j}\aver
{V^\beta(0)\phi_j(1)}_{\mathrm{disk}}^{\mathrm{BCFT}_0^{\mathrm{matter}}},\label{bone}
\end{eqnarray}
where we used (\ref{tad-aux}) and
\begin{eqnarray}
\aver{c\bar c(0) c(1)}_{\rm disk}^{\rm ghost}&=&-1. \label{gh-tad}
\end{eqnarray}
Notice that $\aver {V^\beta(0)\phi_j(1)}$ is the basic
bulk-boundary two-point function of the matter part of BCFT$_{0}$.
Using this we can express the coefficients in front of the
Ishibashi states $\Ishibashi{V_\beta}$ in the $\ket{B_\Psi}$
boundary state in terms of the bare bones
\begin{equation}
n^\beta_\Psi=-\frac\pi2\sum_j4^{h_j}\, A_\Psi^{\beta j}\,\aver
{V^\beta(0)\phi_j(1)}_{\mathrm{disk}}^{\mathrm{BCFT}_0^{\mathrm{matter}}}.\label{bone-disk}
\end{equation}
As a consistency check we derive the same formula directly on the
upper half plane
\begin{eqnarray}
-4\pi i\aver{E[\tilde \V^\beta]\Big|c\phi_j}^{\mathrm{BCFT}_0'}&=&-4\pi i\aver{\tilde V^\beta(i,-i)f_I\circ c\phi_j(0)}_{\mathrm{UHP}}^{\mathrm{BCFT}_0'}\nonumber\\
&=&-4\pi i\,\left(f'_I(0)\right)^{h_j-1} \aver{\tilde V^\beta(i,-i) c\phi_j(0)}_{\mathrm{UHP}}^{\mathrm{BCFT}_0'},\\
f_I(z)&=&\frac{2z}{1-z^2}.
\end{eqnarray}
Factorizing the correlator in ghost, matter and the auxiliary
sectors
\begin{equation}
\aver{\tilde V^\beta(i,-i)
c\phi_j(0)}_{\mathrm{UHP}}^{\mathrm{BCFT}_0'}=\aver
{c(i)c(-i)c(0)}_{\mathrm{UHP}}\,\aver{V^\beta(i,-i)\phi_j(0)}_{\mathrm{UHP}}
\,\aver{w^\beta(i,-i)}_{\mathrm{UHP}},
\end{equation}
we find
\begin{eqnarray}
\aver {c(i)c(-i)c(0)}_{\mathrm{UHP}}&=&2i,\\
\aver{w^\beta(i,-i)}_{\mathrm{UHP}}&=&4^{h_\beta-1}\aver{w^\beta(0,0)}_{\mathrm{disk}}=4^{h_\beta-1},
\end{eqnarray}
where use of (\ref{tad-aux})  has been made. In total we thus get
\begin{equation}
\boxed{\phantom{\Biggl(}~ n^\beta_\Psi=-\frac\pi2\sum_j
2^{h_j}4^{h_\beta}\, A_\Psi^{\beta j}\,\aver
{V^\beta(i,-i)\phi_j(0)}_{\mathrm{UHP}}^{\mathrm{BCFT}_0^{\mathrm{matter}}}.\label{bone-UHP}
~~}
\end{equation}
Consistently we find
\begin{eqnarray}
\aver {V^\beta(i,-i)\phi_j(0)}_{\mathrm{UHP}}&=&2^{h_j}\,4^{-h_\beta}\aver {V^\beta(0)\phi_j(1)}_{\mathrm{disk}}\nonumber\\
&=&|f'(0)|^{h_j}|f'(i)|^{2h_\beta} \aver {V^\beta(0)\phi_j(1)}_{\mathrm{disk}}\nonumber\\
f(z)&=&\frac{1+i z}{1-i z}.\nonumber
\end{eqnarray}
Notice that the formulas (\ref{bone-disk}) or (\ref{bone-UHP})
explicitly express the boundary state in terms of the BCFT$_0$
primaries that are switched on in the solution $\Psi$. As we will
see later on, this  is very useful for identification of the
boundary conditions described by any liftable  numerical solution
which, level by level, can always be put to the form
(\ref{lev-sol}). 

\section{Analytic solutions: Rolling tachyon}
\label{s-rt}

The aim of this section is to illustrate our construction in an
explicit case where  Ellwood conjecture has been verified, and all
OSFT computations have been done already. We select the simplest
well-defined OSFT solutions corresponding to marginal deformations
of the initial BCFT$_0$, where the marginal current has regular
OPE with itself. The whole construction can be readily extended
\cite{Ellwood} to the Kiermaier-Okawa solutions \cite{ KO,
Fuchs:2007yy}, as well as to any other example in which the
Ellwood invariant has been shown  to analytically compute the
tadpole shift, for example \cite{BMT, multibranes}. For
definiteness we select the rolling tachyon marginal deformation
generated by the marginal current $V=e^{X^0}$.

These solutions have been  constructed in the ${\cal B}_0$-gauge in \cite{marg1,
marg2} and extended to more general gauges in \cite{Erler:2007rh,Kiermaier:2010cf}
\begin{equation}\label{rolling}
\Psi_\lambda = F c \frac{B}{1+\lambda  e^{X^0}
\frac{1-F^2}{K}} \lambda c e^{X^0} F,
\end{equation}
where $F=F(K)$\footnote{We assume the conditions $F(0)=1$,
$F'(0)<0$ and $F(\infty)=0$.} and $K, B, c$ are the familiar
string fields \cite{Schnabl, Okawa, Erler1, Erler2, lightning},
and $e^{X^0}$ is the insertion of the exactly marginal boundary
operator $:e^{X^0}\!:(s)$ in the sliver frame.

Given an on-shell weight-zero primary closed string state ${\cal
V}=c\bar c V^{(1,1)}$,  the Ellwood invariant for this class of
solutions has
 been computed in three different ways  \cite{Kishimoto:2008zj, Noumi:2011kn, Erler:2012qr},
and the result (with the BCFT$_0$ contribution---given by the tachyon vacuum invariant---conveniently subtracted) is
\begin{eqnarray}
\bra{E[\V]}\Psi-\Psi_{TV}\rangle&=&-\left\langle e^{-\lambda\int_0^{1}d s\;
e^{X^0}(s)}\,
\V(i\infty)c(0)\right\rangle_{C_1}^{\mathrm{BCFT}_0}\nonumber\\
&=&-\frac1{2\pi i}\left\langle e^{-\lambda\int_0^{2\pi}d\theta\;
e^{X^0}\left(e^{i\theta}\right)}\,
\V(0)c(1)\right\rangle_{\mathrm{disk}}^{\mathrm{BCFT}_0}.
\end{eqnarray}
The nontrivial rearrangement of the $e^{X^0}$ insertions in the solution into a simple boundary interaction is a general consequence of the particular
form of the solution and the string field $F(K)$, as discussed in \cite{ Erler:2012qr, Erler:2012qn}.

This closed string tadpole is in fact closely related to the proper overlap of a closed string of the form $c\bar c V_{m}$ with the boundary
state of $\Psi$
\begin{eqnarray}
\left\langle e^{-\lambda\int_0^{2\pi}d\theta\;
e^{X^0}\left(e^{i\theta}\right)}\,
c\bar c V_m^{(h,h)}(0)c(1)\right\rangle_{\mathrm{disk}}^{\mathrm{BCFT}_0}&=&\frac12\left\langle e^{-\lambda\int_0^{2\pi}d\theta\;
e^{X^0}\left(e^{i\theta}\right)}\,(c_0-\bar c_0)\,
c\bar c V_m^{(h,h)}(0)\right\rangle_{\mathrm{disk}}^{\mathrm{BCFT}_0}\nonumber\\&\equiv&\frac12 \bra{B_\Psi}c_0^-\ket{c\bar c V_m^{(h,h)}},\label{tad-bound}
\end{eqnarray}
where in the last line we have used the  defining expression for
the boundary state, in particular in the ghost sector we used
(\ref{bcBS-defexpr}). Notice, that although  this relation is
trivially true for any matter operator, there is no gauge
invariant observable in the OSFT defined on BCFT$_0$ (with generic
boundary conditions) that could give the LHS of (\ref{tad-bound})
for $h\neq1$. To overcome this difficulty we lift the solution to
the OSFT based on \be {\rm BCFT}_0'={\rm BCFT}_0\otimes {\rm
BCFT}^{\rm aux}. \ee

Because of the geometric nature of the solution (\ref{rolling}),
the simplest lifting we can do is to replace the BCFT$_0$
worldsheet generated by $K$  with the BCFT$_0'$ one, generated by
$K+K^{\rm aux}$,i.e.\footnote{We could have equivalently selected a
space-like direction $Y$ (along which no boundary primaries are
excited)
 and change its boundary conditions to Dirichlet. }
\begin{equation}
\tilde\Psi=\Psi{\Big|}_{K\to K+K^{\rm aux}}.
\end{equation}
Consider  now  $ \V^{(h)}=c\bar c  V^{(h,h)}$, where $ V^{(h,h)}$
is a weight $h$ level-matched primary of $\textrm{BCFT}_0$. The
state can be turned into a weight zero primary
\begin{equation}
\tilde \V^{(h)}=\V^{(h)}e^{2\sqrt{h-1}Y}e^{\frac{2i\phi}{\sqrt3}},
\end{equation}
which now has a nonvanishing tadpole in BCFT$_0'$, chosen as in
section~\ref{ss-gei}.

Now we compute an Ellwood invariant in this slightly modified OSFT
\begin{eqnarray}
\bra{I}\tilde\V^{(h)}(i)|\tilde\Psi-\tilde\Psi_{TV}\rangle&=&-\left\langle
e^{-\lambda\int_0^{1}d s\; e^{X^0}(s)}\,
\tilde\V^{(h)}(i\infty)c(0)\right\rangle_{C_1}^{\mathrm{BCFT}_0'}\nonumber\\
&=&-\frac1{2\pi i}\left\langle e^{-\lambda\int_0^{2\pi}d\theta\;
e^{X^0}\left(e^{i\theta}\right)}\,
\tilde\V^{(h)}(0)c(1)\right\rangle_{\mathrm{disk}}^{\mathrm{BCFT}_0'}\nonumber\\
&=&-\frac1{2\pi i}\langle c\bar c(0) c(1)\rangle \left\langle
e^{-\lambda \int_0^{2\pi}d\theta\;
e^{X^0}\left(e^{i\theta}\right)}
   V^{h,h}(0)\right\rangle^{{\mathrm{BCFT}}_0}\left\langle e^{2\sqrt{h-1}Y}e^{\frac{2i\phi}{\sqrt3}}(0)\right\rangle^{\mathrm{BCFT}^{\rm aux}}\nonumber\\
&=&-\frac1{4\pi i}\left\langle e^{-\lambda \int_0^{2\pi}d\theta\;
e^{X^0}\left(e^{i\theta}\right)}
 (c_0-\bar c_0) c\bar c  V^{h,h}(0)\right\rangle_{\mathrm{disk}}^{\mathrm{BCFT}_0},
\end{eqnarray}
where we have used (\ref{tad-aux}). We thus found
\begin{eqnarray}
\bra{c\bar c  V^{(h,h)}}c_0^-\ket{B_\Psi}&=&-4\pi i\,\bra{I}\tilde\V^{(h)}(i)|\tilde\Psi-\tilde\Psi_{TV}\rangle\nonumber\\
&=&\left\langle c\bar c V^{(h,h)}(0)
 (c_0-\bar c_0) e^{-\lambda \int_0^{2\pi}d\theta\;
 e^{X^0}\left(e^{i\theta}\right)}\right\rangle_{\mathrm{disk}}^{\mathrm{BCFT}_0}.
\end{eqnarray}

Once this is true for any level-matched primary of
${\textrm{CFT}}^{\rm matter}$, it follows from the Virasoro gluing
conditions that

\begin{equation}
\ket{B_\Psi}=e^{-\lambda \int_0^{2\pi}d\theta\; e^{X^0}}\ket{B_0},
\end{equation}
where $\ket{B_0}$ is the boundary state of
$\textrm{BCFT}_0$\footnote{In general, such a formula formally
defines the boundary state for any kind of perturbation. When the
operators inserted on the boundary do not commute, path ordering
is needed.} .

 To elucidate the relation between Ishibashi states and Ellwood invariants
 we  can compute the energy momentum tensor of the solution. Following Sen (appendix A of \cite{sen-review}), the energy momentum tensor can be extracted from
the general form of a boundary state describing a configuration of
branes in flat Minkowski spacetime
\begin{equation}\label{Sen-bound}
\ket B=\int \frac{d^{26}k}{(2\pi)^{26}}\left[ F(k)+( A_{\mu\nu}(k)+ C_{\mu\nu}(k))\alpha_{-1}^\mu\bar\alpha_{-1}^\nu
+ B(k)(b_{-1}\bar c_{-1}+\bar b_{-1}c_{-1})+...\right]c_0^+ c_1\bar c_1\ket{0,k},
\end{equation}
where
\begin{eqnarray}
A_{\mu\nu}&=&A_{\nu\mu} ,\\
C_{\mu\nu}&=&-C_{\nu\mu}.
\end{eqnarray}
Using this in the linearized equation of motion of Closed String Field Theory, one finds that the source of the graviton (i.e. the energy-momentum tensor)
is given by
\begin{equation}
 T_{\mu\nu}(k)=\frac12\left( A_{\mu\nu}(k)+\eta_{\mu\nu} B(k)\right).\label{stress}
\end{equation}
By inspection we find that\footnote{We use the normalization for
the BPZ inner product \bdm\aver{0,k|c_{-1}c_0c_1\bar c_{-1}\bar
c_0\bar c_1|0,k'}=(2\pi)^{26}\delta(k+k').\edm}
\begin{eqnarray}
 A^{\mu\nu}(k)&=&-\frac1{2}\bra{0,-k}c_{-1}\bar c_{-1}\alpha_1^{(\mu}\bar\alpha_1^{\nu)}\, c_0^{-} \ket{B}, \\
 B(k)&=& \frac1{2}\bra{0,-k}c_{-1}\bar c_{-1}\,\frac12(c_1\bar b_1+\bar c_1 b_1)\, c_0^{-}
 \ket{B}.
\end{eqnarray}
 Notice that
$B(k)$ is the overlap of the boundary state with the ghost
dilaton
$$(c\partial^2c - \bar c \bar\partial^2 \bar c)e^{ik\cdot X},$$
which is not a primary field. This seems to imply that $B(k)$
cannot be computed from an Ellwood invariant. However, using the
$bc$-gluing conditions
\begin{equation}
(c_1+\bar c_{-1})\ket B=(b_1-\bar b_{-1})\ket B=0,
\end{equation}
we find that $B(k)$ is also the overlap with the closed string tachyon, which is a primary field
\begin{equation}
 B(k)=\frac1{2}\bra{0,-k}c_{-1}\bar c_{-1}\, c_0^{-} \ket{B}.
\end{equation}
In other words, looking at (\ref{Sen-bound}), we find, on general grounds
\begin{equation}
 B(k)=- F(k).
\end{equation}
Since we are studying a spatially homogeneous process, only the
timelike component $k_0\equiv- iq$ of the momentum enters the
computation.\footnote{We define $\ket{e^{q X^0}}\equiv\ket{0,-i
q}=\ket{0,k_0}$ and we mimic the needed Wick rotation in time by
setting $\bra{ e^{q X^0}}e^{q'
X^0}\rangle^{X^0}=2\pi\delta(q+q')$.
 }
We have
\begin{eqnarray}
 B_\Psi(q)&=&\frac1{2}\bra{e^{-q X^0}}c_{-1}\bar c_{-1}\, c_0^{-}
 \ket{B_\Psi}=-2\pi i \, \bra{E[{\tilde\V}_T]}\tilde\Psi-\tilde\Psi_{TV}\rangle,\\
{\tilde\V}_T&=&c\bar c\, e^{-q X^0}\,e^{2\sqrt{\frac{q^2}{4}-1}\,Y}e^{\frac{2i\phi}{\sqrt3}}.
\end{eqnarray}
From the previous computation we find
\begin{eqnarray}
B_\Psi(q)&=&-f_\lambda(q)\;\mathrm{Vol}_{25},\\
f_\lambda(q)&\equiv&\left\langle e^{-\lambda
\int_0^{2\pi}d\theta\; e^{X^0}\left(e^{i\theta}\right)}\,e^{-q
X^0}(0)\right\rangle_{\mathrm{disk}}^{X^0},
\end{eqnarray}
 Notice  that
$B_\Psi(q)$ is
just the coefficient of the Ishibashi state $\Ishibashi{e^{q X^0}}$   in the boundary state.
Continuing with $A^{ij}_\Psi(q)$ we have
\begin{eqnarray}
A^{ij}_\Psi(q)&=&-\frac1{2}\bra{e^{-q X^0}}c_{-1}\bar
c_{-1}\alpha_1^{(i}\bar\alpha_1^{j)}
\, c_0^{-} \ket{B_\Psi}=2\pi i \, \bra{E[{\tilde\V^{ij}}]}\tilde\Psi-\tilde\Psi_{TV}\rangle,\\
\tilde\V^{ij}&=&-2c\bar c\partial X^{(i}\bar\partial X^{j)} e^{-q X^0}\,e^{q Y}e^{\frac{2i\phi}{\sqrt3}}.
\end{eqnarray}
Computing the Ellwood invariant gives
\begin{equation}
A^{ij}_\Psi(q)=-f_\lambda(q)\,\delta^{ij}\,\mathrm{Vol}_{25},
\end{equation}
and trivially also
\begin{equation}
A^{i0}_\Psi(q)=0.
\end{equation}
It is less straightforward to compute $A^{00}_\Psi$. To get this
contribution we have to contract the boundary state with the
closed string state  $W=-2 \no{\partial X^0\bar\partial X^0 e^{-q
X^0}}$. Due to normal ordering this is not a primary field (for
nonzero momentum) and we cannot directly compute this contribution
from the Ellwood invariant. We have to first decompose the state
in primaries and descendants,
 and use the Virasoro gluing conditions for the boundary state to
 reexpress the contribution from descendants in terms of primaries
\begin{eqnarray}
\bra{e^{-q X^0}}c_{-1}\bar
c_{-1}\alpha^0_1\bar\alpha^0_1=-\frac2{q^2}\bra{e^{-q X^0}}
c_{-1}\bar c_{-1}L_1^{\mathrm{matter}}\bar
L_1^{\mathrm{matter}},\quad\quad q\neq0.
\end{eqnarray}
For nonzero momentum we thus have
\begin{eqnarray}
A^{00}_\Psi(q)&=&-\frac1{2}\bra{e^{-q X^0}}c_{-1}\bar c_{-1}\alpha_1^{0}\bar\alpha_1^{0}\, c_0^{-} \ket{B_\Psi}\nonumber\\
&=&\frac1{2}\frac2{q^2}\bra{e^{-q X^0}}c_{-1}\bar c_{-1}L_1^{\mathrm{matter}}\bar L_1^{\mathrm{matter}}c_0^-\ket{B_\Psi}\nonumber\\
&=&\frac1{q^2}\bra{e^{-q X^0}}c_{-1}\bar c_{-1}[L_1^{\mathrm{matter}}, L_{-1}^{\mathrm{matter}}]c_0^-\ket{B_\Psi}\nonumber\\
&=&\frac1{q^2}\,2 \frac{q^2}{4}\bra{e^{-q X^0}}c_{-1}\bar c_{-1}c_0^-\ket{B_\Psi}\nonumber\\
&=&B_\Psi(q)=-f_\lambda(q)\,\mathrm{Vol}_{25},\quad\quad
q\neq0.\label{A00}
\end{eqnarray}
Notice that this contribution comes from the first nontrivial
level of the Ishibashi state
$$\Ishibashi{e^{q X^0}}=\left(1+\frac1{2h}L^{\mathrm{matter}}_{-1}\bar L^{\mathrm{matter}}_{-1}+...\right)\ket{e^{q X^0}},$$
thus it is not surprising that we get the same result as if we contracted the boundary state with the closed string tachyon.

For $q=0$, $\partial X^0\bar\partial X^0 e^{-q X^0}$ is a primary
 and probes the Ishibashi state $\Ishibashi{\partial X^0\bar\partial
X^0}$. Thus for zero momentum we have
\begin{eqnarray}
A^{00}(q=0)&=&-\frac1{2}\bra{0}c_{-1}\bar c_{-1}\alpha_1^{0}\bar\alpha_1^{0}\, c_0^{-} \ket{B_\Psi}=2\pi i\, \bra{E[{\V^{00}}]}\Psi-\Psi_{TV}\rangle,\label{335}\\
\V^{00}&=&-2c\bar c\partial X^0\bar\partial X^0,
\end{eqnarray}
which gives
\begin{eqnarray}
A^{00}(q=0)&=&\mathrm{Vol}_{25}\left\langle e^{-\lambda \int_0^{2\pi}d\theta\;
 e^{X^0}\left(e^{i\theta}\right)}\,(-2)\partial X^0\bar\partial X^0(0)\right\rangle_{\mathrm{disk}}^{X^0}\nonumber\\
&=&\mathrm{Vol}_{26}.
\end{eqnarray}
Notice that because of momentum conservation the boundary
interaction is  not giving any contribution, this would not be the
case for the $\cosh X^0$ deformation. The energy momentum tensor
in the $q$-space is thus given by
\begin{eqnarray}
T^{ij}_\Psi(q)&=&\frac12\left(A_\Psi^{ij}(q)+\delta^{ij}B_\Psi(q)\right)=-f_\lambda(q)\,\delta^{ij}\, \mathrm{Vol}_{25},\\
T^{i0}_\Psi(q)&=&0,\\
T^{00}_\Psi(q)&=&\frac12\left(A_\Psi^{00}(q)+\eta^{00}B_\Psi(q)\right)=0,\,\quad q\neq0,\\
T^{00}_\Psi(q=0)&=&\frac12\left(A_\Psi^{00}( 0)+\eta^{00}B_\Psi(0)\right)=\frac12\left[\mathrm{Vol}_{26}+f_\lambda(0)\mathrm{Vol}_{25}\right].
\end{eqnarray}
Looking at the definition of $f_\lambda$ we see that
\begin{equation}
f_\lambda(0)=\langle e^{-\lambda\int
e^{X^0}}\rangle_{\mathrm{disk}}^{X^0}=\langle1\rangle_{\mathrm{disk}}^{X^0}=\mathrm{Vol}_{X^0},
\end{equation}
since our zero-mode normalization is such that
\begin{equation}
\mathrm{Vol}_{X^0}=\bra 0 0\rangle=2\pi \delta(0).
\end{equation}
So in total we find
\begin{equation}
T^{00}_\Psi(q)=2\pi\delta(q)\,\mathrm{Vol}_{25}.
\end{equation}
It remains to compute the disk amplitude $f_\lambda(q)$. In fact,
this  amplitude has been computed by Larsen {\it et al.} in
\cite{Larsen}. Here our approach is seemingly different but
equivalent: instead of getting time dependence by isolating the
time zero mode from the path integral, we contract with the state
$e^{-q X^0}$ and then Laplace transform in $q$ to the `closed
string time' $x^0$, as we discuss later. Combining the results of \cite{Larsen} with the
appropriate momentum conservation, we get
\begin{eqnarray}
f_\lambda(q)&=&\left\langle e^{-\lambda \int_0^{2\pi}d\theta\; e^{X^0}\left(e^{i\theta}\right)}\,e^{-q X^0}(0)\right\rangle_{\mathrm{disk}}^{X^0}\\
&=&\sum_{n=0}^{\infty} (-2\pi\lambda)^n 2\pi\delta(n-q),
\end{eqnarray}
where we took advantage of the disk geometry and the fact that the
distance between the boundary insertions and the bulk insertion is
always 1.

To find the explicit dependence in time we should in principle
Wick rotate,  Fourier transform and Wick-rotate back. A
tailor-made shortcut for this particular example is just to
Laplace transform in real time, without any Wick rotation. In
particular we have
\begin{equation}
f_\lambda(x^0)\equiv\int_0^\infty \frac{dq}{2\pi}\,f_\lambda(q)e^{q X^0}=\frac1{1+2\pi\lambda\, e^{x^0}}.
\end{equation}
We then find
\begin{eqnarray}
\frac{T^{ij}(x^0)}{\mathrm{Vol}_{25}}&=&-\frac1{1+2\pi\lambda\, e^{x^0}}\delta^{ij}\,,\\
\frac {T^{00}(x^0)}{\mathrm{Vol}_{25}}&=&1.
\end{eqnarray}

This is the usual  energy-momentum tensor for a half S-brane
exhibiting  energy conservation and exponential decay for the
pressure.\footnote{ We noticed the following curiosity: if we
change the boundary condition on $X^0$ to Dirichlet
$X^0(0,\pi)=x^0$ then (\ref{rolling}) is no more a solution,
because the boundary operator $e^{X^0}=e^{x^0}$ is now a weight
zero {\it number}. This off-shell string field is however a state
in the $KBc$ algebra and we can easily compute its Ellwood
invariant with a graviton vertex operator $\V^{ij}=-2c\bar c \partial
X^i\bar\partial X^j$ in quite full generality \cite{multibranes}, to
find
$$
2\pi
i\aver{E[\V^{ij}]|\Psi_\lambda-\Psi_{TV}}^{(X^0\to\mathrm{Dir})}=-\frac{1}{1+w\lambda
e^{x^0}}\delta^{ij}\mathrm{Vol}_{25}= T^{ij}_{\lambda\to
\frac{w\lambda}{2\pi}}(x^0)=T^{ij}_\lambda(x^0+\log \frac w{2\pi})
$$
where $w\equiv-\frac{d}{dK}F^2(K){\big|}_{K=0}>0$. Note that
although the invariant  depends now on the choice of security
strip (the `solution' is no more a solution so changing the
security strip is no more a gauge transformation), the dependence
is physically irrelevant as it can be absorbed by a redefinition of the
marginal parameter and thus removed by a shift in time. This is reminiscent of
previous observations  on the  late time behavior of the rolling tachyon
solutions \cite{ell-roll,Hellerman:2008wp}.}

\newpage

\section{Numerical solutions: Lumps in Siegel gauge}
\label{s-lumps}

The aim of this section is to show how to construct the boundary
state for numerical solutions in the level expansion. Our
interest is in the Siegel-gauge lump solutions initially studied
in \cite{MSZ, Moeller, Beccaria, thesis} and recently constructed
to greater accuracy by two of us (M.K, M.S.) \cite{KS}.

\subsection{Moeller--Sen--Zwiebach lump at $R=\sqrt3$}

The first examples of lower D-branes appearing in string field via
inhomogenous tachyon condensation are the tachyon lump solutions
found by Moeller, Sen and Zwiebach \cite{MSZ}. They construct lump
solutions along a compact direction $X$ with radius $R$. Imposing
Siegel gauge, twist symmetry and spatial symmetry under
reflections $X \to -X$, such solutions are given up to level $L=3$
in terms of the towers
\begin{eqnarray}
\ket{T_n}&=&c_1\,\cos\left(\frac n R X(0)\right)\ket0,\nonumber\\
\ket{U_n}&=&c_{-1}\,\cos\left(\frac n R X(0)\right)\ket0,\nonumber\\
\ket{V_n}&=&c_{1}L^{(X)}_{-2}\,\cos\left(\frac n R X(0)\right)\ket0,\\
\ket{W_n}&=&c_{1}L'_{-2}\,\cos\left(\frac n R X(0)\right)\ket0,\nonumber\\
\ket{Z_n}&=&c_1L_{-1}^{(X)}L_{-1}^{(X)}\,\cos\left(\frac n R X(0)\right)\ket0\nonumber,
\end{eqnarray}
in the form
\begin{equation}
\ket\Psi=\sum_{n|L\leq 3}(t_n\ket{T_n}+u_n\ket{U_n}+v_n\ket{V_n}+w_n\ket{W_n}+z_n\ket{Z_n}).\label{lump}
\end{equation}
The Virasoro generators appearing in the expansion of the solution are purely matter, and are split according to the decomposition of the energy momentum tensor
$$
T^{\mathrm{matter}}_{c=26}(z)=T^{(X)}_{c=1}(z)+T'_{c=25}(z),
$$
in the two BCFT sectors
$$ \mathrm{BCFT}^{\mathrm{matter}}_{c=26}=\mathrm {BCFT}^{X}_{c=1}\otimes \mathrm {BCFT}'_{c=25}.$$
Ghost degrees of freedom are spanned by ghost oscillators. More
zero momentum primaries of BCFT$^{X}$ appear at higher  levels and
a more convenient basis is thus given by oscillators in the
$X$-direction, \cite{Beccaria, KS}. For the time being we consider
lump solution of the form (\ref{lump}) at radius $R=\sqrt{3}$. For
this particular value, the reader can  find the numerical results
for the lump coefficients $(t_n,u_n,v_n,w_n,z_n)$  in table 3 of
\cite{MSZ}.

Our aim is to use the result we derived in section~\ref{ss-EIBP}, which allows us to define the boundary state in terms of the primaries
that are switched on in the solution. The formulas (\ref{bone-disk}) and (\ref{bone-UHP}) provide a linear expression for the coefficients of the Ishibashi states in terms of
 the coefficients of the solution.
 We will be interested especially in computing the energy density profile  of
the lump and its pressure along the direction $X$ on which the
lump is forming. These quantities can be easily obtained from
generalized Ellwood invariants.

 What is needed is a lift for the
numerical level-truncated solution. Since the solution is not
turning on any primary along $\mathrm {BCFT}'_{c=25}$, a simple
lift is given by (\ref{lump}), with the replacement \be L'\to
L'+L^{\rm aux}. \ee Equivalently, instead of tensoring with an
auxiliary BCFT of $c=0$, we can just impose the Dirichlet boundary
condition on an arbitrary space direction in $\mathrm
{BCFT}'_{c=25}$, say $Y\equiv X^{25}$, along which the solution
does not change. Because of the universal structure in the
$Y$-direction, the solution remains a solution and the
coefficients of the Ishibashi states we compute are not affected
by the new Dirichlet boundary conditions.


For the energy profile we have to compute the following generalized Ellwood invariants
\begin{equation}
E_n\equiv -4\pi i\aver{E[c\bar c \partial X^0\bar\partial X^0\,e^{i\frac{nX}{R}+\frac{nY}{R}}]\,{\Big|} \Psi-\Psi_{TV}}.
\end{equation}
The $n=0$ contribution is precisely the mass of the brane configuration,
normalized to 1 for a single lower dimensional D-brane; it is the coefficient of the
Ishibashi state of the zero momentum graviton in the time-time
direction. In terms of the momenta $E_n$, the energy density
profile can be defined as a simple Fourier series\footnote{This is
$T^{00}=\frac12(A^{00}-B)$ as defined in (\ref{stress}). In the
present context, it is easy to check that $A^{00}=-B=E$, see
(\ref{335}). In this section we set the volume of CFT$'$ to unity.
Moreover, we normalize the zero mode in the $X$-CFT by setting
$\bra0 0\rangle=R$. With this choice the observables we compute
are naturally related to integer numbers.}
\begin{equation}
E(x)\equiv T^{00}(x)=\frac1{\pi R}\left(\frac12 E_0+\sum_{n=1}^\infty E_n \cos\frac{nx}{R}\right).\label{gauge-profile}
\end{equation}
If a solution describes a lower dimensional brane sitting at $x=0$, its energy density profile should be given by
\begin{equation}
E(x)=\delta(x)=\frac1{\pi R}\left(\frac12 +\sum_{n=1}^\infty  \cos\frac{nx}{R}\right).
\end{equation}
Thus an exact lump solution sitting at $x=0$ will be characterized by
\begin{equation}
E_n=1,\quad \forall n=0,...,\infty \qquad{\textrm {(Exact Lump)}}.
\end{equation}

To compute the pressure we need in addition  the coefficient of the Ishibashi state for a zero momentum graviton along
the $X$-direction, which is captured by the Ellwood invariant
\begin{equation}
D\equiv 4\pi i\aver{E[c\bar c \partial X\bar\partial X]\,{\Big|} \Psi-\Psi_{TV}}.
\end{equation}
This quantity measures how much the original Neumann boundary
conditions  on the $X$-BCFT are changed to Dirichlet by the
solution. If the solution describes the perturbative vacuum with Neumann boundary conditions
we will have
\begin{equation}
D^{N}=+R\sim\aver{\partial X\bar\partial X(0)}_{\mathrm{disk}}^{\mathrm{Neumann}},
\end{equation}
while for a single lump we should have
\begin{equation}
D^{D}=-1\sim\aver{\partial X\bar\partial X(0)}_{\mathrm{disk}}^{\mathrm{Dirichlet}}.
\end{equation}
While the non-constant Fourier modes of the pressure transverse to
the lump are trivially zero,\footnote{The pressure itself is
defined as $T^{XX}=\frac12(A^{XX}+B)=\frac12(A^{XX}-A^{00})$, see
the discussion around (\ref{stress}). Its computation proceeds
analogously to the $T^{00}$ computed in the previous section for
the rolling tachyon, see (\ref{A00}). The nonzero momentum part of
$A^{XX}$ is computed by writing $\partial X\bar\partial X e^{i
nX/R }$ as a descendant of the tachyon $e^{i nX/R }$, which
precisely cancel the corresponding contribution from $B=-A^{00}$.
So only the zero momentum part is nontrivial and it gives rise to
the expression (\ref{lump-pressure}).} its constant mode is given
by
\begin{equation}
P\equiv-\frac12\left(D+E_0\right).\label{lump-pressure}
\end{equation}
For the perturbative vacuum, both quantities add up with the same sign giving $P=-E_0$, but for the exact lump solution they should cancel each other
\begin{equation}
P=0 \qquad{\textrm {(Exact Lump)}},
\end{equation}
i.e. the pressure transverse to a D-brane is zero.

Let's see how to
compute the $E_n$'s and $D$.  In order to do so, it is very
convenient to use the conservation laws for the Ellwood invariant.
These conservation laws have been first derived in \cite{KKT} and
\cite{Kishimoto:2008zj}. In appendix~\ref{a-cons} we offer more streamlined derivation
and we use it to write down few more such laws that are essential for the study of lumps.
The gauge invariant combinations $E_n$ and $D$ can be obtained from Ellwood invariants associated to weight-zero closed-string primary vertex operators
$${\cal V}^{h}=c\bar c V_{\mathrm{CFT'}}^{(1-h)}\,V_{\mathrm{CFT}_X}^{(h)}$$
with weight $(h,h)$ in the $X$-sector.
Applying the conservation laws we find
\begin{eqnarray}\label{conser}
\bra {E[{\cal V}^{h}]}U_n\rangle&=&\bra {E[{\cal V}^{h}]}T_n\rangle,\nonumber\\
\bra {E[{\cal V}^{h}]}V_n\rangle&=&-\left(16 h-\frac12\right)\,\bra {E[{\cal V}^{h}]}T_n\rangle,\nonumber\\
\bra {E[{\cal V}^{h}]}W_n\rangle&=&\left(16 h-\frac72\right)\,\bra {E[{\cal V}^{h}]}T_n\rangle,\nonumber\\
\bra {E[{\cal V}^{h}]}Z_n\rangle&=&-2\frac{n^2}{R^2}\,\bra {E[{\cal V}^{h}]}T_n\rangle.
\end{eqnarray}
Thus, up to level 3
\begin{eqnarray}
E_n&=&-4\pi i\aver{E[c\bar c \partial X^0\bar\partial X^0\,e^{i\frac{nX}{R}+\frac{nY}{R}}]\,{\Big|} \Psi-\Psi_{TV}}
\nonumber\\
   &=&-4\pi i\sum_{m|L\leq 3} f_{nm}\;\aver{E[c\bar c \partial X^0\bar\partial X^0\,e^{i\frac{nX}{R}+\frac{nY}{R}}]\,{\Big|} T_m},
\end{eqnarray}
where the coefficients $f_{nm}$ are given by the conservation laws (\ref{conser})
\begin{equation}
f_{nm}=-\delta_{m0}\frac 2\pi +t_m +u_m -\left(4\frac{n^2}{R^2}-\frac12\right)v_m+\left(4\frac{n^2}{R^2}-\frac72\right)w_m-\frac{2m^2}{R^2}z_m.
\end{equation}
Notice that the tachyon vacuum has been subtracted from the
solution by the $-\frac2\pi \ket{T_0}$ term in $f_{nm}$. We are
left with a single Ellwood invariant whose computation gives, see
(\ref{bone-UHP}),
\begin{equation}
-4\pi i \aver{E[c\bar c \partial X^0\bar\partial
X^0\,e^{i\frac{nX}{R}+\frac{nY}{R}}]\,{\Big|} T_m}= -\frac{\pi R}
2\,4^{\frac{n^2}{R^2}}\,\delta_{nm}\, \frac{1+\delta_{n0}}{2}.
\end{equation}
The last factor came from the overlap of the cosine mode with the exponential momentum mode in the closed string vertex operator.
In the end we thus find
\begin{equation}
E_n=-\frac{\pi R} 2\,4^{\frac{n^2}{R^2}}\, \frac{1+\delta_{n0}}{2}\,f_{nn}.
\end{equation}

\begin{figure}
 \begin{minipage}[b]{0.5\linewidth}
 \centering
 \resizebox{3.2in}{1.7in}{\includegraphics[scale=1]{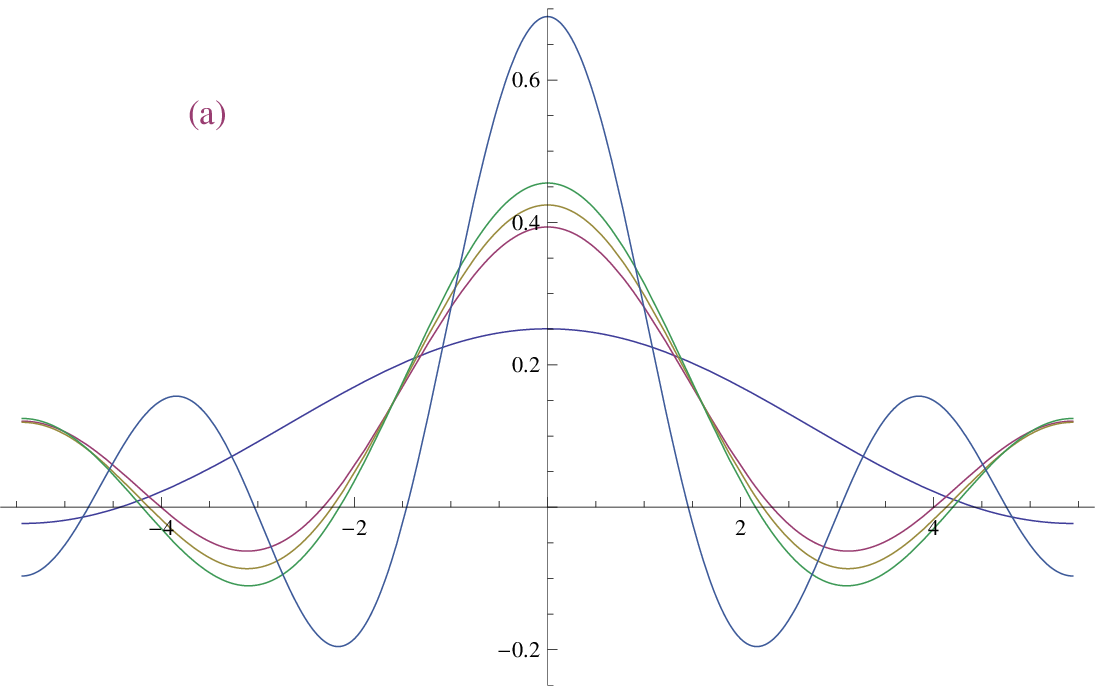}}
 \label{fig:figure1}
 \end{minipage}
 \hspace{0.5cm}
 \begin{minipage}[b]{0.5\linewidth}
 \centering
 \resizebox{3.2in}{1.7in}{\includegraphics[scale=1]{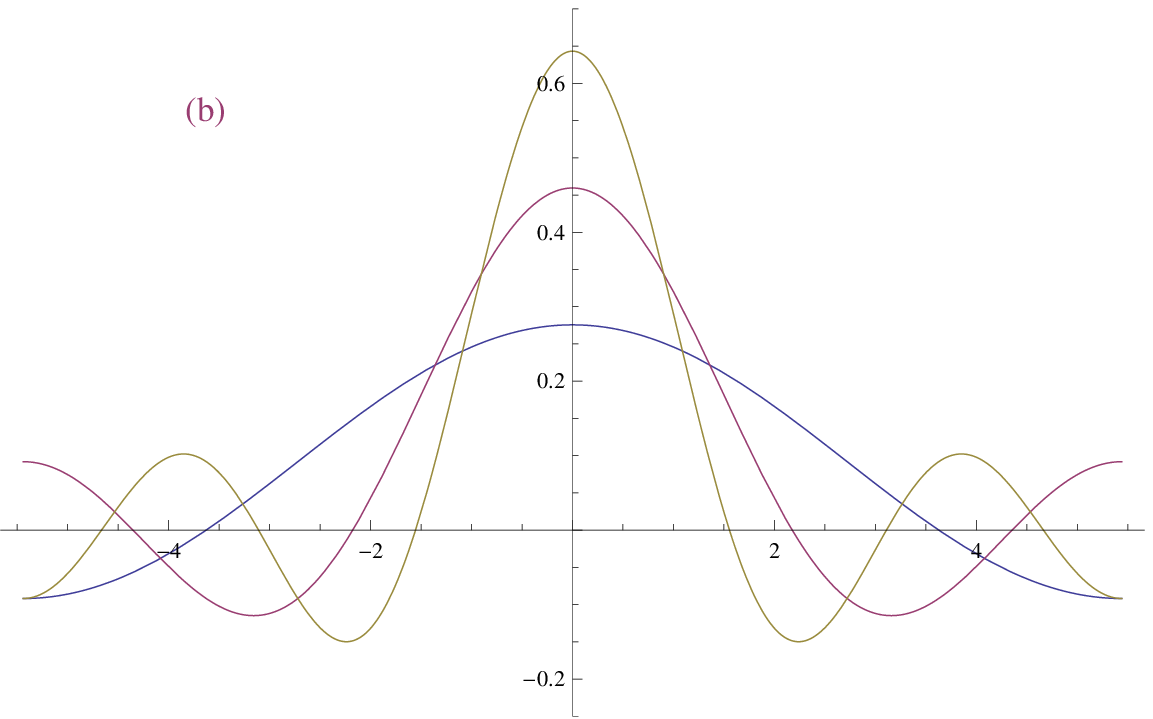}}
 \label{fig:figure2}
 \end{minipage}
 \caption{\label{fig:profile}{\small$(a)$ Gauge-invariant energy density profile of the Siegel-gauge-lump at $R=\sqrt3$,
at levels $L=\frac1 3,\,\frac4 3,\,2,\,\frac7 3,\, 3$, as defined
by eq. (\ref{gauge-profile}). At level 1/3 only one harmonic is available and corresponds to the less localized profile.
 At levels $L=4/3,\, 2,\, 7/3$ the second
harmonic enters the game and the profile is essentially unchanged
till  $L=3$ where the third harmonic gives a substantial
contribution. $(b)$ Plot of $\frac1{\pi
R}\left(\frac12+\sum_{n=1}^N \cos \frac nR x\right)$, for
$N=1,2,3$, at $R=\sqrt 3$. This is how the delta-function forms in
an expansion in harmonics.}}
 \end{figure}

\begin{figure}
\begin{center}
\resizebox{3.5in}{1.4in}{\includegraphics{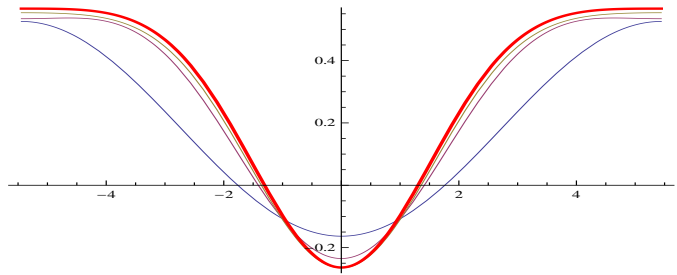}}
\end{center}
\caption{\label{fig:profile2} Traditional open string tachyon profile of the Siegel-gauge-lump at
$R=\sqrt 3$, at levels $L=\frac1 3$ (blue line), $2$ (magenta line), $3$ (yellow line) and $L=12$ (red thick line):  higher harmonics are
suppressed in the Fock space and the open string profile is essentially unchanged as we increase the level.}
\end{figure}

The computation of $D$ proceeds analogously but in a simpler way, since only the zero momentum part of
the solution participates. Using the conservation laws (\ref{conser}) we find
\begin{eqnarray}
D&=& 4\pi i\aver{E[c\bar c \partial X\bar\partial X]\,{\Big|} \Psi-\Psi_{TV}}\nonumber\\
&=&4\pi i\, d_0\aver{E[c\bar c \partial X\bar\partial X]\,{\Big|} T_0}\nonumber\\
&=&-\frac{\pi R}{2}\,d_0,
\end{eqnarray}
where, up to $L=3$
\begin{equation}
d_0=-\frac2\pi+t_0+u_0-\frac{31}2 v_0+\frac{25}2 w_0.
\end{equation}
Using the coefficients given in table 3 of \cite{MSZ} we find the following values
\begin{center}
\begin{tabular}{|c|c|cccc|c|c|}
\hline
$L$ $(R=\sqrt3)$&${\mathrm{Action}}$   & $E_0$      & $E_{1}$  &   $E_{2}$ &    $E_{3}$ &     $D$  &      $P$         \\
\hline
$1/3$  &         $1.32002$  &        $1.23951$  & $0.743681$  & $-       $ &  $-       $ & $1.23951$ &  $-1.23951 $   \\
\hline
$4/3$  &         $1.25373$  &        $1.14776$  & $0.741903$  & $0.825738$  & $-       $  & $1.14776$  &$-1.14776 $   \\
\hline
$2$      &       $1.11278$  &        $1.10298$  & $0.830459$  & $0.927894$  & $-       $  & $-0.574734$  &$-0.264122$   \\
\hline
$7/3$  &         $1.07358$  &        $1.07489$  & $0.899585$  & $1.0405$ &    $-       $ & $-0.992768$  &$-0.0410632 $ \\
\hline
$3$        &     $1.06421$  &        $1.0645$  & $0.89973$  &   $1.07981$ &   $1.23776 $ & $-1.08289$  &$0.00919196 $ \\
\hline
Expected        &     $1$  &        $1$  & $1$  &   $1$ &   $1 $ & $-1$  &$0 $ \\
\hline
\end{tabular}
\end{center}
In the first column we have also written down the mass of the lump as computed from the action, see column $r^{(1)}$ in table 4 of \cite{MSZ}.
The pressure is nicely going to zero.
To give an optical visualization we plot the energy density profile in figure \ref{fig:profile}a. To compare
 we plot the approximants of the delta function $\frac1{\pi R}\left(\frac12+\sum_{n=1}^N \cos \frac nR x\right)$ for $N=1,2,3$,
 see figure \ref{fig:profile}b.
It is also interesting to qualitatively compare with the known
open-string-tachyon profile (given by $\sum_n t_n \,\cos \frac n R
x$, see figure \ref{fig:profile2}). It is apparent that in the
`closed-string' profile of figure \ref{fig:profile} the higher
harmonics play an essential role in localizing it to zero width,
while this does not happen in the open string profile. This is a
consequence of the geometry of  the identity string field,
 which effectively dresses the tachyon coefficients  $t_n$ with $4^{\frac{n^2}{R^2}}$ thus amplifying the effect of higher harmonics.
As it often happens, subleading contributions in the Fock space can have important sizable effects in observables.

Up to here, we have just used the coefficients given in
\cite{MSZ}. There, the maximum level reached was $L=3$, which allowed us to prove our assertions about the energy profile and the Ellwood invariants with few percent accuracy. Obviously, better accuracy is always desirable, but more fundamentally, one could worry that subtleties related to the identity string field might start manifesting themselves at higher levels.
With the code developed by two of the authors
\cite{KS} it is possible to go up to $L=10$ with a reasonable
personal computer power and to explore the lump solutions for
different radii. Using cluster facilities we arrived to level 12
 and, for $R=\sqrt3$ in the same $(L,2L)$ scheme as \cite{MSZ}, we
found\footnote{At higher levels it is more convenient to span the
state space of BCFT$^{X}$ with oscillators acting on momentum
modes. The conservation laws that are needed to compute the above
invariants are derived in appendix~\ref{a-cons}.}

{\footnotesize
\begin{center}
\begin{tabular}{|c|c|ccccccc|c|}
\hline
$L$   & ${\mathrm{Action}}$   & $E_0$      & $E_{1}$  &   $E_{2}$ &    $E_{3}$ &     $E_4$  &                  $E_5$       &$E_6$                   &$D$       \\
\hline
$ 1$  &         $ 1.32002$  &        $1.23951$  & $0.74368$  & $-$ &     $-       $ & $- $ &$-$ & $-$&       $1.23951 $   \\
\hline
$ 2$  &         $ 1.11278$  &        $1.10298$  & $0.83046$  & $0.927897$ &     $-       $ & $- $ &$-$ & $-$&       $-0.574733 $   \\
\hline
$ 3$  &         $ 1.06421$  &       $ 1.0645$  & $0.899731$  & $1.07981$  &     $1.23776$  & $-$ &$-$   &$-$     &$-1.08289$   \\
\hline
$4$      &            $1.03731$  &        $1.04598$  & $0.917185$  & $0.940436$  &   $1.3942$  &  $-     $   &    $-$    &$-$    &$-0.841392$   \\
\hline
$5$  &               $1.03006$  &        $1.04024$  & $0.939622$  & $0.935288$ &    $0.7325$ &   $-       $  &    $-$  &$-$   &$-0.782913 $ \\
\hline
$6$        &         $1.02141$  &        $1.02947$  & $0.945748$  & $0.992221$ &    $0.688677 $ & $1.9835$ &     $-$   &$-$   &$-0.932662$ \\
\hline
$7$        &         $1.01893$  &        $1.02694$  & $0.956255$  & $0.995945$ &    $1.08921 $ & $2.07561$ &     $-$   &$-$   &$-0.955753 $ \\
\hline
$8$       &           $1.01477$  &        $1.02304$  & $0.959451$  & $0.977503$ &    $1.12074 $ & $-0.232145$ &   $-$  &$-$   &$-0.912786 $ \\
\hline
$9$       &           $1.01363$  &        $1.02183$  & $0.965378$  & $0.977053$ &    $0.869997 $ & $-0.30867$ &   $3.5736 $  &$-$   &$-0.91416 $ \\
\hline
$10$       &           $1.0112$  &        $1.01792$  & $0.96745$  & $0.993393$ &    $0.86486 $ & $1.9131$&   $ 3.75069$  &$-$   &$-0.961774 $ \\
\hline
$11$       &           $1.01058$  &        $1.01715$  & $0.97137$  & $0.993749$ &    $1.04171 $ & $1.97574$&   $-3.86516$  &$-$   &$-0.9657950$ \\
\hline
$12$       &           $1.008998$  &        $1.01550$  & $0.97278$  & $0.98704$ &    $1.051325 $ & $0.023453$&   $- 4.14698$ &$8.073065$    &$-0.94658 $ \\
\hline
 Exp.   &           $1$  &        $1$  & $1$  & $1$ &    $1 $ & $1$&   $1$ &$1$    &$-1 $ \\
\hline

\end{tabular}
\end{center}
} The first three lines reproduce the results obtained before
using the coefficients of \cite{MSZ}. All the energy harmonics
appearing at $L=3$ show better approximation to the correct value
1. However starting at level 6 the $E_4$ harmonic enters the game
and it will take some more levels for it to start converging to 1.
Indeed it appears that higher harmonics oscillate quite
erratically before converging to the expected value.\footnote{The
convergence of the $E_n$'s in level truncation does not appear to
be uniform and the level at which $E_n$ starts converging
increases with $n$. Technically speaking, the energy profile must
be understood as
$$ E(x)\equiv\frac1{\pi R}\lim_{N\to\infty} \left[\lim_{L\to\infty}\left(\frac12 E_0^{(L)}+\sum_{n=1}^{N} E_n^{(L)}\cos\frac{nx}{R}\right)\right].$$
In terms of the geometrical definition of the level expansion this means that one would need to consider
$$\lim_{z\to1}\lim_{L\to\infty}\bra{E[V]}z^{L_0}\ket{\Psi^{(L)}},$$
so  that we first let the approximate solution $\Psi^{(L)}$ to
converge to the exact solution and then we send the regulating
strip to zero width. In this way there is a natural cutoff given
by $z^{\frac{n^2}{R^2}}$ for higher harmonics. } We notice that
also the invariant $D$ (which is a genuine Ellwood invariant) is
oscillating with a not so clear pattern with a behavior similar
to the one observed in \cite{japan}.

\subsection{Double lumps at $R=2\sqrt3$}
\label{ss-R2sq3}

For $R>2$  multiple lump solutions are energetically reachable
from the perturbative vacuum via tachyon condensation. As a
further application of our formalism we considered double lump
solutions. These belong to the family of recently discovered
numerical solutions in \cite{KS}. The interest here is to show how
our gauge invariant expression for the energy density can be used
to measure the distance between the lower dimensional branes
described by the solution. Another gauge invariant measurement of
the distance would be given by the mass of stretched strings
between the multiple separated D-branes, which might be harder to
measure in the level expansion with enough precision, as it would
require a careful study of the linearized fluctuations.

\begin{figure}
\begin{center}
\resizebox{3.5in}{1.6in}{\includegraphics{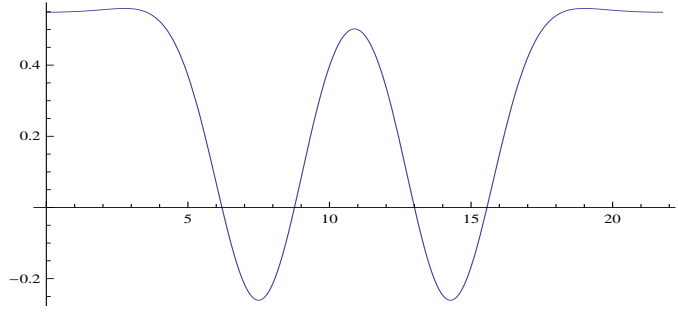}}
\end{center}
\caption{\label{fig:2lumps-open} {\small Open string tachyon profile of a two-lump solution obtained at $R=2\sqrt3$ and level $L=(12,36)$.}}
\end{figure}

Suppose we have a solution $\Psi_a$ describing two D-branes on a circle of radius $R$, symmetric around the origin and at a
distance $ a\,(2\pi R)$ from each other. The energy of the solution will be given by
\begin{equation}
E_0=2,
\end{equation}
meaning that we have two lower dimensional branes.
But how does the number $a$ show up in the $E_n$'s? The exact profile of a double lump configuration with separation $a\,(2\pi R)$, centered around $\pi R$, is given by
\begin{equation}
E_{(a)}(x)=\delta{\Big(}x-\pi R(1-a){\Big)}+\delta{\Big(}x-\pi R(1+a) {\Big)}=\frac1{\pi R}\left(
\frac12 E_0+\sum_{n=1}^\infty E_n\cos \frac{nx}{R}\right).\label{exact-profile}
\end{equation}
Integrating both sides against $\cos \frac xR$ gives
\begin{equation}
\int_0^{2\pi R}dx\, \cos \left(\frac xR\right) \,E_{(a)}(x)=-2\cos (\pi a)=E_1.
\end{equation}
Thus, in the case of a two-lump solution, the invariant $E_1$ measures the distance between the two D-branes
\begin{equation}
a_1=\frac1\pi\arccos\left(-\frac{E_1}{2}\right).\label{dist-1}
\end{equation}
The arc-cosine is defined here in the standard branch $\arccos(0)= \frac\pi2$. The other branches would give the lengths of all the
possible open strings stretching between the branes and wrapping the circle at the same time.
Higher harmonics  can also be used to compute the distance, and integrating (\ref{exact-profile}) against $\cos \frac {nx}R$ we find
\begin{eqnarray}
E_n=2(-1)^n\cos(n\pi a). \label{Ena}
\end{eqnarray}
Solving this equation for $a$ requires some care in choosing the correct branch of the arc-cosine. This must be done in such a way that the distance computed
from any $E_n$ gives the same value $a_1$ as computed from $E_1$. The result can be written as
\begin{equation}
a_n=(-1)^{p_n}\frac1{\pi n} \arccos \left((-1)^n \frac{E_n}{2}\right)+\frac{ 2\left[\frac{p_n+1}{2}\right]}{n},\quad\quad n>1\label{dist-n}
\end{equation}
where $[x]$ stands for integer part and the integer $p_n$  is uniquely chosen such that $$\frac{p_n}{n}<a_1<\frac{p_{n}+1}{n},$$ which gives
\begin{equation}
p_n=[n a_1].
\end{equation}
Clearly for the  exact solution $\Psi_a$ we should have
\begin{equation}
a_n=a\equiv \mathrm{Distance},\quad\forall n\geq1,
\end{equation}
which is a
quite nontrivial constraint between the various $E_n$, which will
be only approximatively satisfied at finite level.
 For generic multiple lump solutions, the relative distances between
the various D-branes can be computed from the $E_n$ invariants along similar lines.
Let us look at a particular example. At level $(12,\,36)$ we selected a double lump solution
obtained at $R=2\sqrt3$ which displays the open string tachyon profile shown in figure \ref{fig:2lumps-open}.
The gauge invariant data of the solution are given by\footnote{The
solutions at level 2 and 4 (marked in the table with an asterisk)
are actually complex. We show the real part of the observables
(which would contain a tiny imaginary part of order $10^{-5}$ --
$10^{-1}$ depending on the observable and the level). }

{\small
\begin{center}
\begin{tabular}{|c|c|c|cccccc|}
\hline
$L$&Action& $D$&        $E_0$ &     $E_1$ &    $E_2$ &   $E_3$ &    $E_4$                    &    $E_5$              \\
\hline
1    &2.57014   &$2.4209$       &2.4209 & -- 0.816955 & -- 0.54184 & 1.3133 & --     &    --                  \\
\hline
$2^*$    &2.21165   &$-1.69337$       &2.1897 & -- 0.848747 & -- 0.60583 & 1.89707 & -- 1.62092      &    --                   \\
\hline
3   &2.19355    &$-2.50001$       &2.11767 & -- 0.908501 & -- 0.838798 & 1.84278 & -- 1.24372   &-- 0.987367               \\
\hline
$4^*$   & 2.06874  &$-1.39183$     &2.08709 & -- 0.919667 & -- 0.850043 & 1.88425 & -- 1.0523      & -- 1.02488                \\
\hline
5  & 2.05531    &$-1.37542$     &2.07382 & -- 0.983959 & -- 0.812633 & 1.91245 & -- 1.15202     & -- 0.57724                \\
\hline
6   &2.03894     &$-2.09185$    &2.05368 & -- 1.00138 & -- 0.788653 & 1.92175 & -- 1.30591      &-- 0.518028                \\
\hline
7  &2.03494     &$-2.1419$     &2.04912 & -- 1.03283 & -- 0.765547 & 1.90846 & -- 1.35827       &-- 0.488344                  \\
\hline
8& 2.0269       &$-1.71527$    &2.04119 & -- 1.04599 & -- 0.743696 & 1.90879 & -- 1.35485       & -- 0.42022                   \\
\hline
9& 2.02525       &$-1.70495$     &2.03899 & -- 1.06273 & -- 0.734362 & 1.91644 & -- 1.37781     & -- 0.37505                   \\
\hline
10& 2.02052      &$-2.07063$     &2.03154 & -- 1.07229 & -- 0.717661 & 1.91526 & -- 1.44161     &-- 0.329759                   \\
\hline
11& 2.01969      &$-2.08504$     &2.03029 & -- 1.08369 & -- 0.709787 & 1.90937 & -- 1.45664     &-- 0.295048                   \\
\hline
12&2.01658 &-- 1.81655&2.02687 &-- 1.09091&-- 0.696749&1.90744&-- 1.45907        & -- 0.256288    \\
\hline
Expected& 2&-- 2&2&-- 1.18&-- 0.61&1.90&-- 1.63 & 0.03\\
\hline
\end{tabular}
\end{center}

\begin{center}
\begin{tabular}{|c|ccccccc|}
\hline
$L$   &   $E_6$ & $E_7$ & $E_8$ & $E_9$ & $E_{10}$ & $E_{11}$ &$E_{12}$\\
\hline
1     & --      & --  & --  & --  & --  & -- &--  \\
\hline
$2^*$     & --      & --  & --  & --  & --  & -- &--  \\
\hline
3     & 2.63667 & --  & --  & --  & --  & -- &--  \\
\hline
$4^*$     & 2.80995 & --  & --  & --  & --  & --  &-- \\
\hline
5     & 1.3382 & -- 2.40998 & --  & --  & --  & -- &--  \\
\hline
6     & 1.32239 & -- 2.61623 & -- 0.587783 & --  & --  & -- &--  \\
\hline
7     & 2.08367 & -- 0.516486 & -- 0.299094 & 4.62486 & --  & -- &--  \\
\hline
8     & 2.07383 & -- 0.514651 & -- 0.0446295 & 4.54688 & --  & --  &-- \\
\hline
9     & 1.60158 & -- 2.29289 & -- 0.074498 & -- 2.51466 & -- 6.95806 & --  &-- \\
\hline
10    & 1.58451 & -- 2.37617 & 0.281907 & -- 2.48429 & -- 7.22557 & -- &--  \\
\hline
11    & 1.8759 & -- 1.0672 & 0.380377 & 5.38459 & 8.09111 & 3.54317 &-- \\
\hline
12    &1.86166 &-- 1.07601&-- 0.0969718&5.28465&8.37884&4.10828&    9.35208             \\
\hline
Expected&1.60 &-- 1.92&0.67&1.13&-- 2.00&1.23&0.55                 \\
\hline
\end{tabular}
\end{center}     }
The $(E_0,D)$ invariants and the action are clearly indicating
that we are indeed dealing with a two-lump solution.
The expected values for the $E_n$'s have been derived using
(\ref{Ena}) from the  distance $a_*=0.299\pm0.001$ computed later
on. In appendix~\ref{a-lumps}  another, different two-lump
solution and a single-lump solution are shown for the same value
of $R=2\sqrt3$; all results have been pushed to level $L=(12,36)$.

\begin{figure}
\begin{center}
\resizebox{4in}{2.8in}{\includegraphics{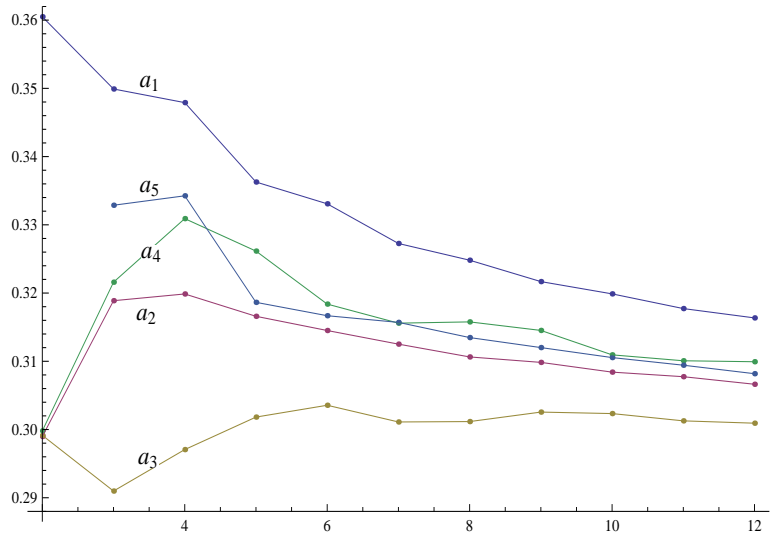}}
\end{center}
\caption{\small{\label{fig:distances} Gauge invariant distances $a_n^{(L)}$ as computed from the first five $E_n$-invariants,
as a function of the level $(L=2,3,...,12)$. Notice that starting from $L=4$ the $a_n$'s show a quite clear (but not so fast) convergence pattern.
}}
\end{figure}

\begin{figure}
\begin{center}
\resizebox{3in}{1.8in}{\includegraphics{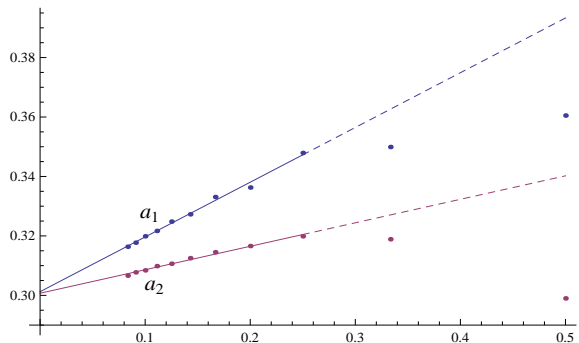}}
\end{center}
\caption{\label{fig:fit-distance}{\small Plot of the  distances $a_{1}$ and $a_{2}$ as functions of $1/L$, together with their best fit
$a^{(\infty,L_{min})}_n+\frac{b_n^{(\infty,L_{min})}}{L}$. The fit here is done in the range $L=4,\ldots,12$,  i.e.  $L_{min}=4$, and it appears to essentially
capture the dependence of the $a_n$'s with the level.}}
\end{figure}

From $E_1$, at the maximal available level $L=12$, we can compute
\begin{equation}
a_1^{(L=12)}=\frac1\pi\arccos\left(-\frac{E^{(12)}_1}{2}\right)=0.316357.
\label{a1}
\end{equation}
Looking at figure \ref{fig:2lumps-open} we see that this is
consistent with the distance  between the minima of the open
string tachyon profile, which, at the same level $L=(12,36)$, is
given by
\begin{equation}
a_{open}^{(L=12)}=0.310439.
\end{equation}
Also $E_2$ and $E_3$ give approximate distances which are quite close to $a_{open}$
\begin{eqnarray}
a_2^{(L=12)}&=&\frac1{2\pi}\arccos\left(\frac{E^{(12)}_2}{2}\right)=0.306633,\\
a_3^{(L=12)}&=&\frac1{3\pi}\arccos\left(-\frac{E^{(12)}_3}{2}\right)=0.300927.
\end{eqnarray}
Going further with the harmonics we have to change the branch of the arc-cosine, according to (\ref{dist-n})
\begin{eqnarray}
a_4^{(L=12)}&=&-\frac1{4\pi}\arccos\left(\frac{E^{(12)}_4}{2}\right)+\frac12=0.309934,\\
a_5^{(L=12)}&=&-\frac1{5\pi}\arccos\left(-\frac{E^{(12)}_5}{2}\right)+\frac25=0.30818,\\
a_6^{(L=12)}&=&-\frac1{6\pi}\arccos\left(\frac{E^{(12)}_6}{2}\right)+\frac13=0.313486.
\end{eqnarray}
Notice, that by (\ref{dist-n}) the $E_n$'s must be bounded by 2 in
absolute value, to be consistent with a two-D-brane
interpretation. The $E_{n\geq6}$'s already show `incorrect values'
up to level 8 and indeed they would need higher level to start
showing a convergence pattern. The set of candidate gauge
invariant distances $(a_1,a_2,a_3,a_4,a_5)$ is plotted in figure
\ref{fig:distances} for the range of levels $L=2,3,...,12$.
To obtain a prediction on the actual distance between the two
D-branes, we have to find a  way to extrapolate the observables of
the approximate level-truncated solution to infinite level. Rather
surprisingly, we observed that a simple $1/L$ fit gives results
which are nicely  consistent within the first five harmonics. For
any harmonic $n=1,...,5$ we fitted the values of $a_n^{(L)}$
\begin{equation}
a_n^{(L)}\approx a^{(\infty,L_{min})}_n+\frac{b_n^{(\infty,L_{min})}}{L},
\end{equation}
in the range of levels $(L_{min},L_{max}=12)$ for all possible $L_{min}$'s in the range
$$4\leq L_{min}\leq (L_{max}-2)=10.$$
The lower bound $4\leq L_{min}$ is justified by figure
\ref{fig:distances} and the upper  bound is  necessary to have at
least 3 points to fit. See figure \ref{fig:fit-distance} for an
example.

By varying $L_{min}$ one can have an estimate of the error of the
linear fit. For any  frequency $n$ we take the mean value $\bar
a_n^{(\infty)}$ from the results obtained for different $L_{min}$
and we take the associated standard deviation $\sigma_{n}$ as a
measure of the error. The obtained values are the following
\begin{center}
\begin{tabular}{|c|c|c|}
\hline
$n$ &$\bar a_n^{(\infty)} $   & $\sigma_n$     \\
\hline
$1$  &         $0.300$  &        $0.001$               \\
\hline
$2$  &         $0.299$  &        $0.001$                 \\
\hline
$3$      &     $0.299$  &        $0.004$               \\
\hline
$4$  &         $0.299$  &        $0.003$             \\
\hline
$5$        &   $0.298$  &        $0.002$             \\
\hline
\end{tabular}
\end{center}
These are five independent `measurements' of the distance and the
fact that they are all  mutually consistent is  quite a nontrivial
check for the linear fit. Given the mutual consistency between
these values, we can average them with weights $w_n=1/\sigma_n^2$
and obtain the value for the distance
\begin{equation}
a_*=0.299\pm0.001,
\end{equation}
where the error has been computed with $\left(\sum_n w_n\right)^{-1/2}$.

\begin{figure}[h]
 \begin{minipage}[b]{0.5\linewidth}
 \centering
 \resizebox{3.2in}{1.7in}{\includegraphics[scale=1]{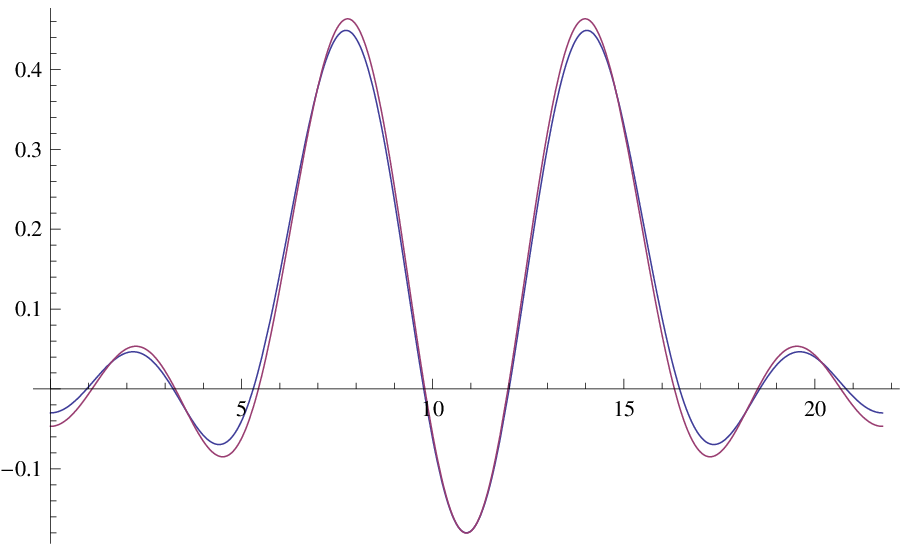}}
 \label{fig:figure1}
 \end{minipage}
 \hspace{0.5cm}
 \begin{minipage}[b]{0.5\linewidth}
 \centering
 \resizebox{3.2in}{1.7in}{\includegraphics[scale=1]{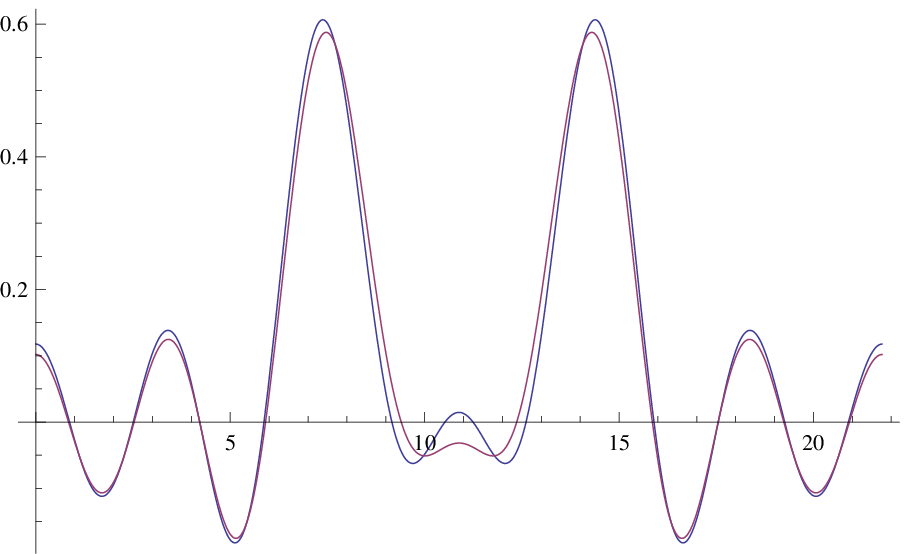}}
 \label{fig:figure2}
 \end{minipage}
 \caption{\label{fig:2lumps}{\small Plot of $\frac1{\pi R}\left(\frac12 E_0+\sum_{n=1}^m E_n \cos\frac{nx}{R}\right)$, in blue line,
 against the corresponding truncation of a sum of two delta
functions, in magenta line, at a separation $\bar a_*=0.3$, for $m=4$ (left) and $m=6$ (right). Here $R=2\sqrt3$ and the coefficients are
obtained at level  $L=(12,36)$. Notice how the profile of the truncated solution displays a slightly bigger effective distance than the `exact' one, as if
the lumps were getting closer by increasing the level.}}
 \end{figure}

Since we are `measuring' a modulus of a BCFT in an unknown point
of its moduli space via an approximate OSFT solution, we do not
have a given value to compare with, but to appreciate to what
extent $a_*\sim 0.3$ is consistent with the distance between the
two D-branes described by the solution, we plot the energy profile
of the solution including up to 6th harmonic against the
corresponding truncation of a sum of two delta functions, at
distance $a=0.3$, see figure \ref{fig:2lumps}.

For completeness, few  more calculations are needed to completely
reconstruct the boundary state for a system of parallel lower
dimensional branes. At zero momentum we have an infinite tower of
bulk primaries with weight $(h^2,h^2)$ for integer $h$. Here we
have only considered the coefficients of the Ishibashi states for
the identity (captured by $E_0$) and for the weight (1,1) primary
$\partial X\bar\partial X$ (captured by $D$). We didn't compute the
coefficients of the Ishibashi states of the higher level zero
momentum primaries.  It would also be necessary to verify that the
coefficients of the winding-mode Ishibashi states are vanishing,
as it must be for Dirichlet boundary conditions. As for the zero
momentum primaries, these coefficients too get contribution only
from the zero momentum part of the solution. We leave these
computations for future work.

\section{Conclusions}
In this paper we have proposed and analyzed a simple shortcut to
compute the BCFT boundary  state corresponding to a  classical
solution of OSFT. In a nutshell, OSFT provides the coefficients
for the linear combination of Ishibashi states forming the
boundary state. These coefficients are given by Ellwood invariants
of a lifted solution in a modified OSFT where a trivial  $\mathrm
{BCFT}^{\mathrm{aux}}$ with $c=0$ has been tensored with the
original BCFT$_0$. We have characterized the lifting procedure somewhat
implicitly, and it would be indeed very interesting to understand it in full generality.
In most cases, however, it is sufficient to
assign Dirichlet boundary conditions to one spacetime direction.
The essence of the trick is to associate to any closed string
primary of the form $c\bar c V^{\mathrm{matter}}$ a corresponding
weight zero primary with nonvanishing tadpole thanks to the
Dirichlet boundary condition,
 without altering the  physics described by the
solution (the solution remains a solution in the tensor theory).
Assuming Ellwood conjecture thus completely defines
 the boundary state.

Given any solution $\Psi$, a family of closed string states $\ket{B_*(\Psi)}^{(s)}$ has been constructed in \cite{KOZ}.
These states are conjectured to be BRST equivalent to the boundary state we deal with in this paper. Among other
 things, the construction depends on a free parameter $s$ and, in the $s\to0$ limit, under some assumptions on the regularity of the solution, one
   recovers the worldsheet geometry of the Ellwood invariant. Our construction,
    where spinless matter primaries are lifted to formal elements
 of the closed string cohomology in an enlarged CFT, applies to \cite{KOZ} as well.
  All the derivations of \cite{KOZ} go through in the tensor theory and, inside the (enlarged)
 closed string cohomology, they clearly agree with our general results of section~\ref{s-bs}, assuming Ellwood
 conjecture.
Thus, in principle, the coefficients of the Ishibashi states
(\ref{KMS}), and hence the full BCFT-boundary state, can be
computed as well from
 \begin{equation}
n_\Psi^\alpha=-\frac1{2}\bra{\tilde {\cal V}^\alpha}(c_0-\bar
c_0)\ket{B_*(\tilde\Psi)}^{(s)}.
 \end{equation}
 The construction at finite $s$ is a gauge invariant deformation and regularization of the generalized Ellwood invariant.
 The final $s$-independence of this deformation is a consequence of closed string linearized gauge invariance (the dependence of
 $\ket{B_*(\tilde\Psi)}^{(s)}$ on $s$ is, in general, at most BRST exact) and, ultimately, of the validity of the OSFT equation of motion,
 \cite{KOZ}. This is an important point, given that
 there are in general infinite families of string fields sharing
 the same Ellwood invariants with OSFT solutions.

Defining the boundary state from OSFT can potentially contribute
to the  development of boundary conformal field theory. The
classification of consistent boundary states in a given CFT
background is, at present, still an open problem. Our proposal
gives a complementary way of determining the coefficients of the
Ishibashi states, without having to deal with complicated
consistency conditions such as Cardy or various sewing conditions
\cite{Gaberdiel}. What we have to do, instead, is to {\it solve}
the OSFT equation of motion. This, at least in principle, is a
clear well defined task and we have at our disposal a lot of
analytic and numerical tools for progressing in this direction.

From the OSFT point of view, it is important to understand whether the
space of classical solutions is bigger or smaller than the space
of consistent boundary conditions. Suppose OSFT has more solutions
than expected: this, for example, could show up in exotic ``non
integer'' values for the
 Ellwood invariants (an example of this possibility has been found \cite{Erler:2010pr}, in the context of cubic super string field theory)
 and hence our boundary state would violate Cardy conditions. On the other hand, our method could prove useful in the search for
  solutions associated with generic BCFT's whose
  boundary state is given.
Matching the generalized Ellwood invariants of a to-be-found solution to a target boundary state with given BCFT moduli, will partially constrain the
coefficients of the solution and  a full
solution can in principle be searched for in the level expansion (or in some other regularization) by extremizing the constrained action.
Consequently one must verify that the full action is  also extremized, along the lines of \cite{Large}.
However it is not guaranteed, see for example \cite{Canadians}, that such a solution would exist for any choice of the BCFT moduli.

\newpage

\section*{Acknowledgments}

\noindent

 We would like to thank Ted Erler, Matthias Gaberdiel, Yuji Okawa and Cornelius Schmidt-Colinet for useful discussions.
 The access to computing and storage facilities owned by parties and
projects contributing to the Czech National Grid Infrastructure
MetaCentrum, provided under the programme "Projects of Large
Infrastructure for Research, Development, and Innovations"
(LM2010005), and to CERIT-SC computing facilities provided under the
programme Center CERIT Scientific Cloud, part of the Operational
Program Research and Development for Innovations, reg. no. CZ.
1.05/3.2.00/08.0144 is highly appreciated. This research was supported by the EURYI grant GACR EYI/07/E010 from EUROHORC and ESF.

\begin{appendix}

\section{Example of $\mathrm {BCFT}^{\mathrm{aux}}$}
\label{a-aux}

An explicit example of $\mathrm {BCFT}^{\mathrm{aux}}$ is given by
tensoring a free boson with Dirichlet boundary condition ($c=1$)
with an appropriate $c=-1$ BCFT. For the first Dirichlet factor it is useful to recall the following one-point
function\footnote{Here, as in the rest of the paper we set $\alpha'=1$. In general one should also include the boundary entropy $g_X$ factor on the right hand side, but for the auxiliary sector such a factor drops out of all the computations in this paper.}
\begin{equation}
 \aver {:e^{q X}:(z,\bar z)}_{\mathrm{disk}}^{\mathrm{Dirichlet}}=(1-|z|^2)^{\frac {q^2}{2}}\label{one-point},
\end{equation}
where we take $q$ to be a generic complex number. For $q$
imaginary, the vertex operator has positive weight. For $q$ real
it is a negative weight primary and it is not normalizable as a
free field. However in the presence of Dirichlet boundary
conditions, the divergence of the zero mode integration is weaker
than the delta function imposed by the boundary condition. To
derive (\ref{one-point}) we start with the Green's function on the
disk with Dirichlet boundary conditions
\begin{equation}
G(z,w)=\aver {X(z,\bar z)X(w,\bar w)}_{\mathrm{disk}}^{\mathrm{Dirichlet}}=-\frac12\log|z-w|^2+\frac12\log|1-\bar z w|^2,\label{Green-Dir}
\end{equation}
where the disk is defined by $|z|\leq 1$.
One can easily check that
\begin{equation}
G(e^{i\theta},w)=G(z,e^{i\theta})=0,
\end{equation}
meaning that the Dirichlet condition $X(|z|=1)=0$ is satisfied.
The first term is just the usual Green's function on the complex
plane (no boundary) while the second is the effect of the
Dirichlet boundary conditions. Only the first (bulk) term in
(\ref{Green-Dir}) should be subtracted for closed string vertex
operators. Then the result (\ref{one-point}) easily follows from
usual Wick contractions.

An example of the needed $c=-1$ theory can be taken to be a free
boson $\varphi$ with background charge $Q$  with an energy
momentum tensor given by
\begin{equation}
T(z)=-:\partial\varphi\partial\varphi:+i Q\partial^2\varphi.
\end{equation}
The central charge is given by
\begin{equation}
c=1-6 Q^2,
\end{equation}
so that we have $c=-1$ for $Q=\frac1{\sqrt{3}}$. The weights of the exponential
primary fields are
\begin{equation}
h[e^{i\alpha\varphi}]=\frac\alpha2\left(\frac\alpha2-Q\right)=\frac\alpha2\left(\frac\alpha2-\frac1{\sqrt3}\right).
\end{equation}
Notice that, differently from the $Q=0$ case, there are now two
operators with vanishing conformal weight, one being the identity operator
and the other being
\begin{equation}
w(z)=e^{\frac{2i}{\sqrt3}\varphi(z)},
\end{equation}
which is in general needed to screen the background charge. The
boundary conditions in the upper half-plane are of Neumann type
\begin{equation}
\varphi(z)=\bar\varphi(\bar z),\quad z=\bar z.
\end{equation}
 We normalize $w$ such that the disk correlator is given by
\begin{eqnarray}
\langle\, w(z)\,\rangle_{\mathrm{disk}}&=&\langle\, \bar w(\bar
z)\,\rangle_{\mathrm{disk}}=\langle
\,w(\xi)\,\rangle_{\mathrm{UHP}}=\langle \,\bar
w(\bar\xi)\,\rangle_{\mathrm{UHP}}=1, \quad\forall z,\xi.
\end{eqnarray}

Another simple option for BCFT$^{\rm aux}$ is to choose a product
of two free bosons ($c=2$) with Dirichlet boundary conditions, and
the symplectic fermion theory with $c=-2$ constructed in
\cite{Kausch:1995py}.\footnote{Further details and references are
summarized in \cite{Gaiotto:2003yb}.} The advantage of such a
choice is that we do not need the analogue of $w$ to saturate the
zero modes on the disk.

\section{Conservation laws  for the Ellwood invariant}
\label{a-cons}

In this appendix we derive a set of useful
conservation laws for the Ellwood invariant. Some of them have
been derived already in \cite{KKT} and \cite{Kishimoto:2008zj}. We
present an alternative simpler derivation.

\subsection{Review of conservation laws of the identity string field}

To start with, it is useful to recall some standard conservation
laws for the identity string field $\ket I$ (see
\cite{cons,wedges} for review). They will later be modified by the
 closed string vertex operators inserted at the midpoint in the Ellwood invariant.
 The first such relation is
the anomalous conservation of $K_n$, for nonvanishing central
charge $c$,
\begin{eqnarray}
 K_n&=&L_n-(-1)^nL_{-n},\\
 \bra I K_{n}&=&\frac{c}{8}n\left(i^n+(-i)^n\right)\,\bra I .\label{K-I-conslaw}
 \end{eqnarray}
Another useful conservation law is given by the modes of
current $i\partial X$, which reads
\begin{eqnarray}
 A_n&=&\alpha_n+(-1)^n\alpha_{-n},\\
 \bra I A_{n}&=&0.\label{A-I-conslaw}
\end{eqnarray}
Because of non-anomalous momentum conservation in the $X$-CFT,
this conservation law shows no anomaly. Let's come to the ghost
sector. From the conservation of $K_n$ we can read-off, by
analogy, the conservation of $B_n$, which is not anomalous ($b(z)$
is a genuine weight two primary)
\begin{eqnarray}
 B_n&=&b_n-(-1)^nb_{-n},\\
 \bra I B_{n}&=&0.\label{B-I-conslaw}
\end{eqnarray}
There is also an analogous (anomalous) conservation law for the $c$-ghost which
reads
\begin{eqnarray}
 C_n&=&c_n+(-1)^nc_{-n},\\
 \bra I C_{2n}&=&-(-1)^n\bra I C_0, \label{ceven-I-conslaw}\\
 \bra I C_{2n+1}&=&-(-1)^n\bra I C_1.\label{codd-I-conslaw}
 \end{eqnarray}

\subsection{Virasoro conservation laws}

In this section we compute general conservation laws for Virasoro
generators.  Suppose that the total CFT is the tensor product
$\rm{CFT}^{(1)}\otimes \rm{CFT}^{(2)}$, of two CFT's  with central
charges $c$ and $-c$ respectively. The energy-momentum tensor
decomposes as
\begin{equation}
T(z)=T^{(1)}(z)+T^{(2)}(z).
\end{equation}
The weight-zero vertex operator entering the Ellwood invariant can
be written as
\begin{equation}
V(z,\bar z)=V^{(h,\bar{h})}_{(1)} V^{(-h,-\bar{h})}_{(2)} (z,\bar{z}),
\end{equation}
where $h$ and $\bar{h}$ are the holomorphic and antiholomorphic weights (not necessarily the same) of $V_{(1)}$ with respect to $T^{(1)}$.
The corresponding BRST and conformally invariant Ellwood state is given by
\begin{equation}
\bra {E[V]}=\bra I V^{(h,\bar{h})}_{(1)} V^{(-h,-\bar{h})}_{(2)} (i,-i).
\end{equation}
We start the computation of the conservation law for the modes of
$T^{(1)}$, using the anomalous derivation
$$K^{(1)}_n=L_n^{(1)}-(-1)^n L_{-n}^{(1)}=\oint\frac{dw}{2\pi i}\, v_n(w)
T^{(1)}(w),$$ where the holomorphic vector field $v_n(w)$ is given
by
\be
v_n(w)=w^{n+1}-(-1)^nw^{-n+1}.\label{vn}
\ee
Assuming that $V^{(h,\bar{h})}_{(1)}$ factorizes into the holomorphic and antiholomorphic parts,
we have
\bea
 \langle E[V]|K^{(1)}_n&=&\bra I V^{(h,\bar{h})}_{(1)}
V^{(-h,-\bar{h})}_{(2)} (i,-i)K^{(1)}_n\label{IKV}\\
&=&\oint_0\frac{dw}{2\pi i}v_n(w)\,\bra I V^{(h)}_{(1)}(i)
V_{(1)}^{(\bar h)}(-i) V^{(-h,-\bar{h})}_{(2)} (i,-i)
T^{(1)}(w).\nonumber
\eea
Now, using the formalism of \cite{Leclair, cons}, we write $\bra
I=\bra0 U_f$, with
$$f(w)=\frac{2w}{1-w^2}$$ being the identity conformal map and we
move the operator $U_f$ to the right of the other operators
\bea
&&\oint_0\frac{dw}{2\pi i}v_n(w)\,\bra I V^{(h)}_{(1)}(i)
V_{(1)}^{(\bar h)}(-i) V^{(-h,-\bar{h})}_{(2)} (i,-i) T^{(1)}(w)\nonumber\\
&=&\oint_0\frac{dw}{2\pi i}v_n(w)\,\bra 0 V^{(h)}_{(1)}(i)
V_{(1)}^{(\bar h)}(-i) \left([f'(w)]^2\,T^{(1)}(f(w))+\frac
{c}{12}S_f(w)\right)V^{(-h,-\bar{h})}_{(2)} (i,-i)U_f\nonumber.
\eea
At this point we notice that the geometry of the identity string
field, together with the involved operator insertions at the
midpoint, implies that
\be
\oint_0=-\frac12\oint_{(i,-i)}.
\ee
To see this, we start with the simple observation that the
integrand has poles only in $(0,\pm i,\infty)$. In particular, note that there is no pole at $\pm 1$. The poles at the
midpoint $\pm i$ arise from the $T-V$ contractions and from  the
Schwarzian derivative
\be
S_f(w)=\{f(w),w\}=\frac6{(1+w^2)^2}.
\ee
Then we notice
\bea
f\left(-\frac1w\right)&=&f(w),\\
f'\left(-\frac1w\right)&=&w^2 f'(w),\\
v_n\left(-\frac1w\right)&=&\frac1{w^2}v_n(w),\\
S_f\left(-\frac1w\right)&=&w^4 S_f(w),
\eea
from which it follows from contour deformation that
$$\oint_0=\oint_\infty=-\frac12\oint_{(i,-i)}.$$
We thus get
\bea
\langle E[V]|K^{(1)}_n&=&-\frac12\oint_{(i,-i)}\frac{dw}{2\pi
i}v_n(w)[f'(w)]^2\,\bra 0 V^{(h)}_{(1)}(i) V_{(1)}^{(\bar
h)}(-i) \,T^{(1)}(f(w))V^{(-h,-\bar{h})}_{(2)}U_f
\nonumber\\
&&-\frac c{24}\oint_{(i,-i)}\frac{dw}{2\pi i}v_n(w)S_f(w)\bra 0
V^{(h)}_{(1)}(i) V_{(1)}^{(\bar h)}(-i)
\,V^{(-h,-\bar{h})}_{(2)}(i,-i)\,U_f
\nonumber \\
&=&-\frac12\oint_{i}\frac{dw}{2\pi i}v_n(w)[f'(w)]^2\,\bra
0\left[\frac{h V_{(1)}^{(h)}(i)}{(f(w)-i)^2}+\frac{\partial
V_{(1)}^{(h)}(i)}{f(w)-i}\right]\bar V_{(1)}^{(\bar h)}(-i)
\,V^{(-h,-\bar{h})}_{(2)}(i,-i)\,U_f
\nonumber\\
&-&\!\frac12\oint_{-i}\frac{dw}{2\pi i}v_n(w)[f'(w)]^2\,\bra 0
V_{(1)}^{( h)}(i)\left[\frac{\bar h V_{(1)}^{(\bar
h)}(-i)}{(f(w)+i)^2}+\frac{\partial V_{(1)}^{(\bar h)}(-i)}{f(w)+i}\right]
 \,V^{(-h,-\bar{h})}_{(2)}(i,-i)\,U_f
 \nonumber\\
&-&\!\frac c{24}\oint_{(i,-i)}\frac{dw}{2\pi i}v_n(w)S_f(w)\bra 0
V^{(h)}_{(1)}(i) V_{(1)}^{(\bar h)}(-i)
\,V^{(-h,-\bar{h})}_{(2)}(i,-i)\,U_f
\nonumber.
\eea
It is easy to see that the terms proportional to the non primary
operators $\partial V$ give vanishing contribution.\footnote{This
would not be the case if we worked in the geometry of the local
coordinate $w$ where we would have ended with the singular
insertion $\sim v_n(i)\bra I \partial V(i)\sim 0\times \infty$. In this
case one would need to displace the insertion a bit away from the
midpoint and send the regulator to zero after taking the residue.
This gives a finite net contribution which adds up to the naive
contribution from the double pole. In the global coordinate
geometry $\tilde w=f(w)$ no regularization is needed and the total
contribution just comes from the ``double pole'' $\sim
\frac1{(f(w)\pm i)^2}$. Similar considerations apply to the other
conservation laws we discuss next.} Everything thus simplifies
down to
\bea
&=&-\frac12\!\oint_{(i,-i)}\frac{dw}{2\pi i}v_n(w)\left[\frac
c{12}S_f(w)+[f'(w)]^2\left(\frac {h}{(f(w)-i)^2}+\frac
{\bar h}{(f(w)+i)^2} \right)\right]\langle E[V]|\nonumber\\
&=&n\left[i^n\left(\frac{c}{8}-4h\right)+(-i)^n\left(\frac{c}{8}-4\bar
h\right)\right]\,\bra{E[V]}.
\eea
Summarizing, we found
\begin{eqnarray}
\boxed{\phantom{\Biggl(}~
 \bra{E[V_{(1)} V_{(2)}]}K^{(1)}_n=n\left[i^n\left(\frac{c}{8}-4h\right)+(-i)^n\left(\frac{c}{8}-4\bar h\right)\right]\,\bra{E[V_{(1)} V_{(2)}]}\,.~~}
\end{eqnarray}
The conservation law for $K^{(2)}_n$ is simply obtained by changing $c\to-c$ and $(h,\bar h)\to(-h,-\bar h)$.

\subsection{Oscillator conservation laws}

It is useful to derive the conservation laws for the current
$i\sqrt{2}\partial X$ of a free boson. We focus on Ellwood states with
two kinds of closed string vertex operators: pure momentum modes
and the zero momentum primary $\partial X\bar \partial X$. First we compute
the conservation laws of $\alpha$ oscillators acting on  momentum
modes. We define
\begin{equation}
A_n=\alpha_n+(-1)^n\alpha_{-n}=\oint\,\frac{dw}{2\pi
i}\,g_n(w)\,i\sqrt2\partial X(z).
\end{equation}
The function $g_n(w)$ is defined as
\be
g_n(w)=w^n+(-1)^nw^{-n},
\ee
and obeys
\be
g_n\left(-\frac1w\right)=g_n(w).\label{g-inv}
\ee
Acting with $A_n$ on an Ellwood state of definite
momentum\footnote{The plane wave $e^{ikX}$ is supplemented with a
primary in a decoupled sector of weight $-\frac{k^2}{4}$, which we
call $V_{(2)}$.}
\begin{eqnarray}
\bra I e^{ikX}(i,-i)V_{(2)}(i,-i)A_n&=&i\sqrt2\oint_0
\frac{dw}{2\pi i}
 g_n(w)\bra I e^{ikX}(i,-i) V_{(2)}(i,-i) \partial X(w)\\
 &=&i\sqrt2\oint_0 \frac{dw}{2\pi i}
 g_n(w)f'(w)\bra 0e^{ikX}(i,-i)\partial X(f(w))V_{(2)}(i,-i) U_f\nonumber.
\end{eqnarray}
As in the previous section, here again we notice that, because of
(\ref{g-inv}), we can substitute
$$\oint_0\to-\frac12\oint_{(i,-i)},$$
which is a general property of the identity conservation laws we
consider (but it would not be true for anomalous currents).
Using the OPE
\be
e^{ikX}(i,-i)\partial X(f(w))\sim
-\frac{ik}{2}\left(\frac1{f(w)-i}+\frac1{f(w)+i}\right)e^{ikX}(i,-i),
\ee
and taking the residues at the midpoints we are  left with the
simple result
\begin{eqnarray}
\boxed{\phantom{\Biggl(}~
 \bra I\, e^{ikX}V_{(2)}(i,-i)A_n=-(i^n+(-i)^n)\sqrt{2}\,k\,\bra I\, e^{ikX}V_{(2)}(i,-i)\,.~~}
\end{eqnarray}

In addition, we need the conservation laws for the Ellwood
invariant given by the zero-momentum graviton vertex operator
$c\bar c\partial X\bar\partial X$
\begin{eqnarray}
 &&\bra I c\bar c(i,-i) \partial X\bar\partial X(i,-i) A_n\\
 &=&i\sqrt2\oint_0 \frac{dw}{2\pi i}
 g_n(w)\bra I  \partial X(i)\partial X(-i) \partial X(w)c\bar c(i,-i)\\
 &=&i\sqrt2\oint_0 \frac{dw}{2\pi i}
 g_n(w)f'(w)\bra 0 \partial X(i)\partial X(-i) \partial X(f(w))c\bar c(i,-i)U_f\nonumber\\
 &=&i\sqrt2\left(-\frac12\right)\oint_{(i,-i)} \frac{dw}{2\pi i}
 g_n(w)f'(w)\bra 0 \partial X(i)\partial X(-i) \partial X(f(w))c\bar
 c(i,-i)U_f\nonumber\\
 &=&-\frac i{\sqrt2}\oint_{(i,-i)} \frac{dw}{2\pi i}
 g_n(w)f'(w)\bra 0 \left(-\frac12\frac{\partial X(-i)}{(f(w)-i)^2}-\frac12\frac{\partial X(i)}{(f(w)+i)^2}\right)c\bar
 c(i,-i)U_f.\nonumber
\end{eqnarray}
Taking the residues at the midpoints we get
\begin{eqnarray}
\boxed{\phantom{\Biggl(}~
 \bra I c\bar c \partial X\bar\partial X(i,-i) A_n=\sqrt{2}n^2(-i)^n\,\bra 0c\bar c(i,-i){\Big(}\partial X(i)-(-1)^n\partial
 X(-i){\Big)}U_f.
 ~~}
\end{eqnarray}
Notice that we cannot take  the $U_f$ operator back the vacuum
because the leftover insertion at the midpoint has overall
negative weight. Applying another $A_m$ we get rid of the $\partial X$
insertion and we get
\begin{eqnarray}
\boxed{\phantom{\Biggl(}~
 \bra I c\bar c\partial X\bar\partial X(i,-i) A_n A_m=-2(nm)^2i^{n+m}((-1)^n+(-1)^m)\,\bra 0c\bar
 c(i,-i)U_f.
 ~~}
\end{eqnarray}
Further applications of $A_n$ give trivially zero.

\subsection{Ghost conservation laws}

The conservation law for $b_n$ oscillators is easily obtained from
\begin{eqnarray}
 \bra I c(i) c(-i) V^{(1,1)}(i,-i) B_n&=&\oint_0 \frac{dw}{2\pi i}
 v_n(w)\bra I c(i) c(-i) V^{(1,1)}(i,-i) b(w)\\
 &=&\oint_0 \frac{dw}{2\pi i}
 v_n(w)[f'(w)]^2\bra 0 c(i) c(-i)b(f(w))
 V^{(1,1)}(i,-i)U_f\nonumber\\
 &=&-\frac12\oint_{(i,-i)} \frac{dw}{2\pi i}
 v_n(w)[f'(w)]^2\bra 0 c(i) c(-i)b(f(w))
 V^{(1,1)}(i,-i)U_f\nonumber,
\end{eqnarray}
where we used $$\oint_0=-\frac12\oint_{(i,-i)},$$ just as we did
in the case of energy-momentum tensor.

Performing the midpoint contractions between $b(f(w))$ and $c(\pm
i)$, no residue is found since $v_n(\pm i)=0$, and we are left simply with
\begin{equation}
\boxed{\phantom{\Biggl(}~
 \bra I c(i) c(-i) V^{(1,1)}(i,-i) B_n=0.
  ~~}
\end{equation}

The $c$ ghost conservation law is just a bit more complicated.
What happens here is that the anomalies in the conservation of
$C_n$ on the identity, (\ref{ceven-I-conslaw}) and
(\ref{codd-I-conslaw}), are killed by the $c\bar c(i,-i)$ from the
closed string insertion, as we are now going to see
\begin{eqnarray}
 \bra I c(i) c(-i) V^{(1,1)}(i,-i) C_n&=&\oint \frac{dw}{2\pi i}
 h_n(w)\bra I c(i) c(-i) V^{(1,1)}(i,-i) c(w)\label{c-count}\\
 &=&\oint_0 \frac{dw}{2\pi i}
 h_n(w)[f'(w)]^{-1}\bra 0 c(i) c(-i)c(f(w))
 V^{(1,1)}(i,-i)U_f\nonumber\\
 &=&-\frac12\oint_{(i,-i)} \frac{dw}{2\pi i}
 h_n(w)[f'(w)]^{-1}\bra 0 c(i) c(-i)c(f(w))
 V^{(1,1)}(i,-i)U_f\nonumber.
\end{eqnarray}
Here the quadratic differential $h_n(w)$ is given by
\be
h_n(w)=w^{-2}\left(w^n+(-1)^nw^{-n}\right),
\ee
and obeys
\be
h_n\left(-\frac1w\right)=w^4 h_n(w).
\ee
Once more, this property allows us to replace
$$\oint_0\to-\frac12\oint_{(i,-i)}$$ in going from the second to
third line of (\ref{c-count}). Computing the residues at the
midpoint we are left with
\begin{eqnarray}
&&\bra I c(i) c(-i) V^{(1,1)}(i,-i) C_n \\&=&-i^{n+1}\bra 0
c(i)c(-i){\Big(}c(i)-(-1)^n
c(-i){\Big)}V^{(1,1)}(i,-i)U_f\nonumber.
\end{eqnarray}
Notice that $C_n$ has been localized to a midpoint insertion in
the global coordinate. It is then killed by the two $c$'s from the
closed string insertion. Thus, differently from the pure identity
string field, the conservation law of the $c$ ghost on the Ellwood
state is not anomalous
\begin{equation}
\boxed{\phantom{\Biggl(}~ \bra I c(i) c(-i) V^{(1,1)}(i,-i) C_n=0.
  ~~}
\end{equation}

\section{General properties of the boundary state}\label{a-bound}
In string theory, the boundary state is  a  ghost-number-three
closed string state and it appears as a source term in the closed string
field theory action via the coupling
$$
\bra B c_0^- \ket\Phi
$$
to the dynamical closed string field $\Phi$ of total ghost
number two. The closed string inner product contains the usual insertion of $c_0^-=c_0-\bar c_0$. In order  to write down a  kinetic term for the closed
string field, it is necessary to assume the level matching
conditions \cite{barton}
\be
L_0^-\ket\Phi=b_0^-\ket\Phi=0,
\ee
where $L_0^-=L_0-\bar L_0$ and $b_0^-=b_0-\bar b_0$.
It does not appear consistent to include non-level-matched closed
string states and thus we must impose also
\be
L_0^-\ket B=b_0^-\ket B=0.
\ee
The boundary state is not just a source term in the closed string
action but it is also a peculiar state which incarnates the
existence of a boundary in CFT$_0$ (on which we define closed
string field theory), which preserves conformal invariance.
Together with the previous conditions this means
\bea
b_0^-\ket{B}&=&0\label{tot1},\\
(L_n^{\mathrm{tot}}-\bar L_{-n}^{\mathrm{tot}})\ket{B}&=&0\label{tot2},\\
(Q_{gh}-3)\ket{B}&=&0,\label{tot3}
 \eea
 where
 \be
Q_{gh}=\oint \frac{dz}{2\pi i} (-\no{bc})(z)+ h.c.
 \ee
 is the total ghost number.
\subsection{Proof of matter ghost factorization}
Commuting (\ref{tot1}) with (\ref{tot2}) we learn that
\bea
(b_n-\bar b_{-n})\ket{B}=0,\quad \forall n.\label{b-gluing}
\eea
The most general state obeying (\ref{b-gluing}) can be written in
normal ordered form as
\be
\ket{B}=f{\Big(}\{b_{-m}\},\{\bar b_{-m}\}, c_0^+,[{\rm
matter}]{\Big)}\, \exp\left(-\sum_{n=1}^\infty \bar b_{-n} c_{-n}
+b_{-n} \bar c_{-n}\right)c_1\bar
c_1\ket{0}_{SL(2,C)},\label{gen-b-gluing}
\ee
where $f$ is a generic function depending on $b$-ghost creation
operators, $c_0^+$, and generic matter operators. To see that this
is the case focus on the dependence on $(b_{-n}, \bar b_{-n},
c_{-n}, \bar c_{-n})$ for fixed $n\geq1$. Then conditions
(\ref{b-gluing}) are equivalent to the differential equations
\bea
(\partial_{c_{-n}}-\bar b_{-n}) B(c_{-n},\bar c_{-n}, b_{-n},\bar b_{-n})&=&0,\\
(\partial_{\bar c_{-n}}- b_{-n}) B(c_{-n},\bar c_{-n}, b_{-n},\bar
b_{-n})&=&0,
\eea
whose generic solution is
\bea
B(c_{-n},\bar c_{-n}, b_{-n},\bar b_{-n})=f(b_{-n},\bar b_{-n})
\exp\left(-\bar b_{-n} c_{-n}-b_{-n}\bar c_{-n}\right).
\eea
Repeating this procedure for every $n\geq 1$ and also for $n=0$,
we end up with (\ref{gen-b-gluing}). Finally, imposing ghost
number three, (\ref{tot3}), we conclude that
\be
f{\Big(}\{b_{-n}\},\{\bar b_{-n}\}, c_0^+,[{\rm
matter}]{\Big)}=c_0^+ g([{\rm matter}]).
\ee
Thus we have showed that a state obeying (\ref{tot1}, \ref{tot2},
\ref{tot3}) is necessarily matter--ghost factorized
\be
\ket{B}=\ket{B}^{\rm matter}\otimes\ket{B_{gh}},
\ee
and the ghost factor $\ket{B_{gh}}\equiv \ket{B_{bc}}$ is the
usual boundary state of the $bc$-BCFT
\bea
\left(b_n-\bar b_{-n}\right)\ket {B_{bc}}&=&0,\label{b-glue}\\
\left(c_n+\bar c_{-n}\right)\ket {B_{bc}}&=&0.\label{c-glue}
\eea
From the total gluing conditions (\ref{tot2}) we then find that
$\ket{B_\Psi}^{\rm matter}$ obeys the standard gluing conditions
of the matter sector
\begin{eqnarray}
\left(L^{\mathrm{matter}}_n-\bar
L^{\mathrm{matter}}_{-n}\right)\ket {B}^{\rm matter}&=&0,
\end{eqnarray}
and from this it is easy to check that  that $\ket{B}$ is also
BRST invariant
\be
(Q+\bar Q)\ket{B}=0.
\ee
Few other universal gluing conditions follow from here. Let's look
at the anomalous gluing of the ghost current
$$j_{gh}(z)=-:bc:(z)=\sum_n j_n z^{-n-1},$$ which, using the
gluing conditions (\ref{b-glue}, \ref{c-glue})  reads
\be
(j_n+\bar j_{-n}-3\delta_{n0})\ket{B}=0.
\ee
Notice that this is consistent with (\ref{tot3}),  since $Q_{gh}=j_0+\bar j_0.$ From
here it also follows that the BRST current
$$
j_{BRST}(z)=\sum_n Q_n z^{-n-1}
$$
 glues
non-anomalously at the boundary. Indeed we have that
\be
Q_n=[Q,j_n],
\ee
and thus from the ghost current gluing condition it follows that
\be
(Q_n+\bar Q_{-n})\ket{B}=0.
\ee
\subsection{Normalization of the ghost boundary state}
\begin{figure}
\begin{center}
{\includegraphics{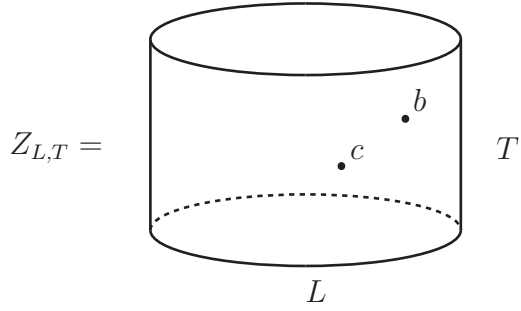}}
\end{center}
\caption{\label{fig:CLT}{\small The partition function $Z_{L,T}$
is given by the path integral on a cylinder $C_{L,T}$, with
insertion of $b$ and $c$. The position of the insertions is
inessential as only the constant zero modes in the expansion of
$b$ and $c$ gives nonvanishing contribution to the path
integral.}}
\end{figure}

Here we fix the normalization of the ghost boundary state from
modular invariance. Consider a cylinder $C_{L,T}$ of circumference $L$ and
height $T$. We put BCFT$_0$ boundary conditions on the
lower and upper boundary of $C_{L,T}$. We are interested in
computing the partition function
$$
\langle 1\rangle_{C_{L,T}}.
$$
In string theory, this partition function is identically vanishing
because the zero modes of the $b,c$ ghosts are not soaked up. On
the cylinder there is a zero mode for $c$ associated with the
constant conformal Killing vector (CKV) for rotation of the
cylinder around its axis. There is also a zero mode for $b$,
associated with the constant holomorphic quadratic differential
(HQD) which changes the length of the base circumference. Because
both the CKV and the HQD are constant we have that
\bea
\langle b(w) c(w')\rangle_{C_{L,T}}&\equiv&Z_{L,T},\\
 \partial_w\partial_{w'}\langle b(w)
c(w')\rangle_{C_{L,T}}&=&0,
\eea
see figure \ref{fig:CLT}.
To compute $Z_{L,T}$  we proceed as follows. We first consider a
cylinder of height $T=\pi$ and circumference $L=2\pi t$. Every
$Z_{L,T}$ can be reduced to $Z_{2\pi t,\pi}$ by simple scaling,
keeping track of the weights of the insertions
\bea
Z_{L,T}=\aver{b
c}_{C_{L,T}}&=&\aver{f\circ b\, f\circ c}_{C_{2\pi
t,\pi}}{\Big |}_{t=\frac L{2T}}=\frac\pi T Z_{2\pi t,\pi}{\Big |}_{t=\frac{L}{2T}}\,,\label{scale}\\
f(w)&=&\frac{\pi w}{T}.\nonumber
\eea

\begin{figure}
\begin{center}
{\includegraphics{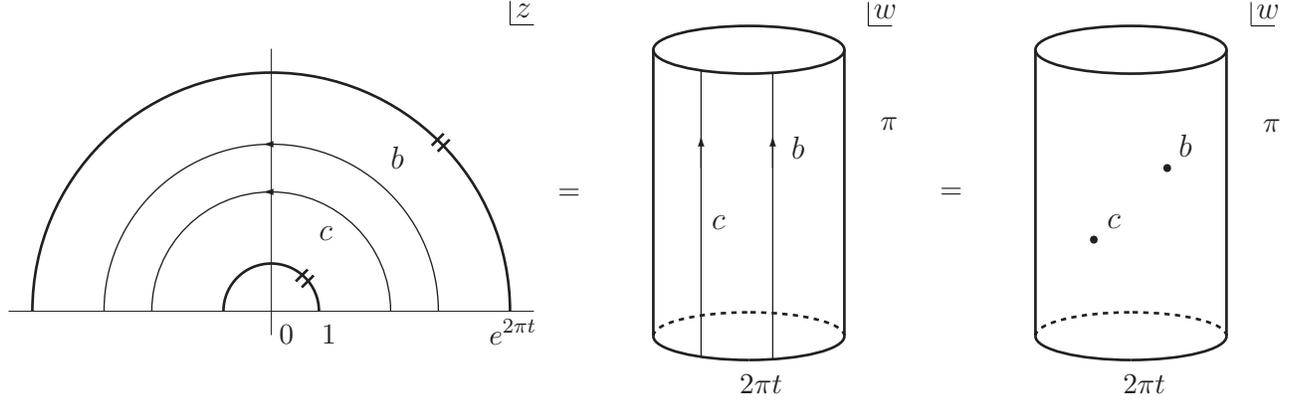}}
\end{center}
\caption{\label{fig:Zopen}{\small Open string trace as a path
integral on the cylinder, as stated in eq.(\ref{oneloop}). The
inner and outer semicircles in the $z$ coordinate are identified
by the $\Tr[(-1)^F(...)]$.}}
\end{figure}

$Z_{2\pi t,\pi}$ is just the one loop open string vacuum amplitude
(before integration over the moduli space)\footnote{We define the
fermion number $$F\equiv Q_{gh}-\frac32=\int_0^{2\pi i}
\frac{dw}{2\pi i}\, j_{gh}(w),$$ as the zero mode of the ghost
current in the canonical strip frame (with doubling trick
understood). $F$ is anti-hermitian and thus $(-1)^F$ is hermitian.
The Siegel-gauge projector $b_0c_0$ is anti-hermitian and the
trace we are computing is imaginary.}
\be
\Tr_{H_{open}}\left[(-1)^F e^{-2\pi t L_0} b_0 c_0\right]=\aver{b
c}_{C_{2\pi t, \pi}}=Z_{2\pi t,\pi}.\label{oneloop}
\ee
This follows from the fact that the cylinder is obtained by
identifying the edges of a canonical open string strip of height
$\pi$ and length $2\pi t$. Such a strip is the image of the half
annulus in the UHP (defined by $1\leq|z|\leq e^{2\pi t}$ and $\Im
z\geq0$), obtained by the map
\bdm
w=\ln z.
\edm
The UHP zero modes $b_0$ and $c_0$ are mapped to vertical line integrals in the $w$
coordinate
\bea
w\circ b_0&=&\oint_0\frac{dz}{2\pi i}\,z\,[w\circ
b(z)]=\frac1{2\pi}\int_0^{2\pi}\, dx\, b(y+i x)\to b(w),\\
w\circ c_0&=&\oint_0\frac{dz}{2\pi i}\,\frac1{z^2}\,[w\circ
c(z)]=\frac1{2\pi}\int_0^{2\pi}\, dx\, c(y+i x)\to c(w),
\eea
where we wrote $w=y+i x$ and extended the UHP by the doubling trick to the full complex plane. In the last step we have `averaged' the integrals because the correlator
only gets contribution from the cylinder zero modes which are
constant in
the $w$ coordinate. This establishes (\ref{oneloop}), see figure \ref{fig:Zopen}\\
We can equivalently compute $Z_{2\pi t,\pi}$ by evolving the
boundary state $\ket {B}$ with the closed string propagator and
contracting with  the  BPZ dual $\bra{B}$. Proceeding
similarly to the open string picture, we can write (using the
ghost gluing conditions)
\bea
\bra{B} e^{-\frac\pi t (L_0+\bar L_0)} (b_0+\bar b_0)(c_0-\bar
c_0)\ket{B}&=&4\bra{B} e^{-\frac\pi t (L_0+\bar L_0)}
b_0c_0\ket{B}\nonumber\\
&=&-4i\, \aver{b c}_{C_{2\pi,\frac\pi t}}\nonumber\\
&=&-4i\,Z_{2\pi,\frac\pi t}.\label{BB}
\eea

\begin{figure}
\begin{center}
{\includegraphics{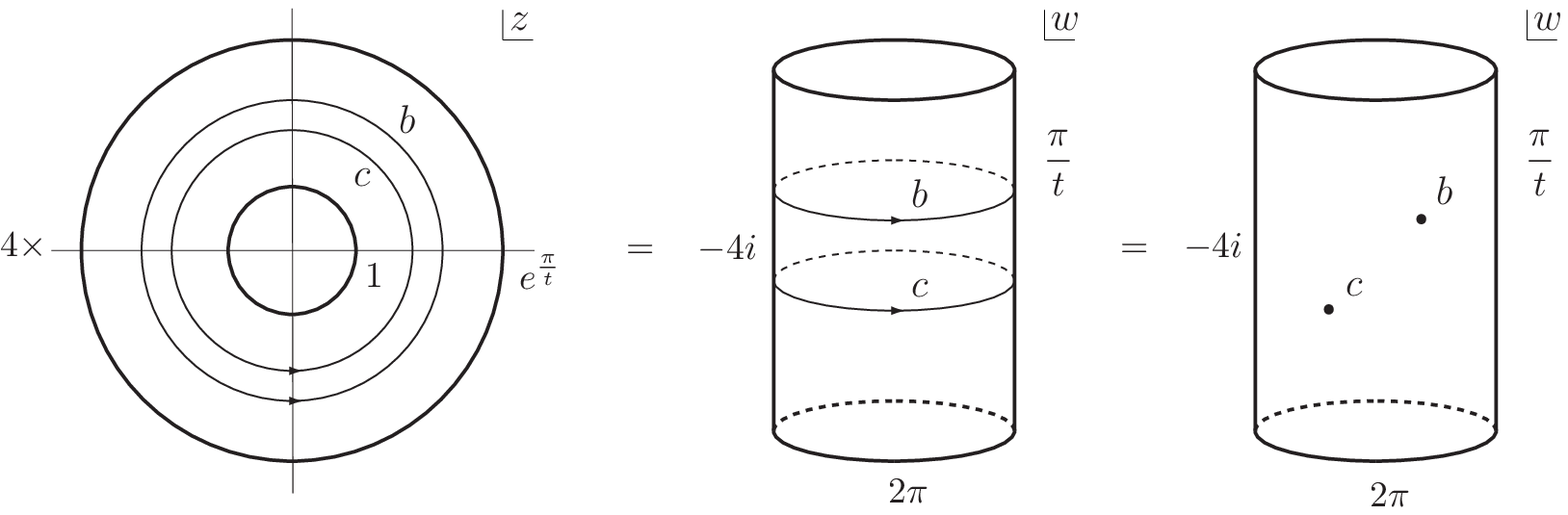}}
\end{center}
\caption{\label{fig:Zclosed}{\small Graphical representation of
eq. (\ref{BB}).}}
\end{figure}
As illustrated in figure \ref{fig:Zclosed}, this is easily
obtained by mapping the annulus $1\leq |z|\leq e^{\frac\pi t}$ to
the cylinder $C_{2\pi,\frac\pi t}$ with
$$ w=i\left(\frac\pi t-\log z\right),$$ and replacing the
resulting {\it horizontal} line integrals with a local insertion
using again
 that the HQD and the CKV for $b$ and $c$  are
constant on the cylinder ($w=y+i x$)
\bea
b_0&\to& \frac1{2\pi}\int_0^{2\pi} dy\, b(y+i x)\to b(w),\\
c_0&\to&-\frac i{2\pi}\int_0^{2\pi} dy\, c(y+ i x)\to -i\, c(w).
\eea
Now we use the scaling property (\ref{scale})
\be
Z_{2\pi,\frac\pi t}=t Z_{2\pi t,\pi},\label{modular-scaling}
\ee
to get
\be
\bra{B} e^{-\frac\pi t (L_0+\bar L_0)} (b_0+\bar b_0)(c_0-\bar
c_0)\ket{B}=-4it\,\Tr_{H_{open}}\left[(-1)^F e^{-2\pi t L_0} b_0
c_0\right].\label{string-cardy}
\ee
This is an equality between two {\it real} quantities. We can use
the above equation to determine the overall normalization of the
ghost boundary state. A generic boundary state for an open string
background is given by
\be
\ket B=\ket {B_{\rm matter}} \otimes \ket {B_{bc}},
\ee
where $\ket {B_{\rm matter}}$ is the matter boundary state
($c=26$) obeying Cardy condition
\be
\bra {B_{\rm matter}}\,e^{-\frac\pi t \left(L^{\rm matter}_0+\bar
L^{\rm matter}_0-\frac c{12}\right)}\ket {B_{\rm
matter}}=\Tr_{H_{\mathrm{open}}^{\rm matter}}\left[e^{-2\pi t
\left(L_0-\frac c{24}\right)}\right],\label{matter-cardy}
\ee
and
\bea
\ket{B_{bc}}&=&{\cal N}_{gh}\,(c_0+\bar
c_0)\exp\left(-\sum_{n=1}^\infty \bar b_{-n} c_{-n} +b_{-n} \bar
c_{-n}\right)c_1\bar
c_1\ket{0}_{SL(2,C)},\label{Bgh}\\
\bra{B_{bc}}&=&-{\cal N}_{gh}\,\bra0 c_{-1}\bar
c_{-1}\exp\left(\sum_{n=1}^\infty \bar b_{n} c_{n} +b_{n} \bar
c_{n}\right)(c_0+\bar c_0),\label{Bgh-bpz}
\eea
are the ghost boundary state and its  BPZ dual whose
normalization we want to determine.
 Taking the ghost part of (\ref{string-cardy}) and assuming
 (\ref{matter-cardy}) we get a Cardy-like condition for the $bc$-BCFT
\be
\bra{B_{bc}}e^{-\frac\pi t \left(L_0+\bar L_0+\frac
{26}{12}\right)} (b_0+\bar b_0)(c_0-\bar
c_0)\ket{B_{bc}}=-4it\,\Tr_{H^{\rm ghost}_{\mathrm{open}}}\left[(-1)^F
e^{-2\pi t \left(L_0+\frac {26}{24}\right)} b_0 c_0\right].
\ee
Computing the left hand side with (\ref{Bgh}, \ref{Bgh-bpz}) we have
\bea
\bra{B_{bc}}e^{-\frac\pi t \left(L_0+\bar L_0+\frac
{26}{12}\right)} (b_0+\bar b_0)(c_0-\bar
c_0)\ket{B_{bc}}&=&-2{\cal N}_{gh}^2 \,\eta^2\!\left(\frac i
t\right)\,\bra{0}c_{-1}\bar c_{-1}(c_0-\bar c_0)(c_0+\bar
c_0)c_1\bar c_1\ket{0}\nonumber\\&=& -2{\cal
N}_{gh}^2\,\eta^2\!\left(\frac i t\right)\, (-2)=4{\cal
N}_{gh}^2\,\eta^2\!\left(\frac i t\right),
\eea
where $$\eta(it)=e^{-\frac{\pi
t}{12}}\prod_{n=1}^\infty\left(1-e^{-2\pi n t}\right)$$ is the
Dedekind $\eta$-function and  we normalize the ghost BPZ-inner
product as\footnote{In the closed string Hilbert space, Hermitian
and BPZ conjugation differ by an overall factor of $i$. In our
conventions (slightly different from \cite{barton})
$${\rm BPZ}\left(\ket0\right)\equiv \bra0=i\; _{hc}\!\bra0\equiv
i\left(\ket0\right)^\dagger.$$ The basic hermitian inner product
is thus given by $$_{hc}\!\aver{0|c_{-1}c_0c_1\bar c_{-1}\bar c_0
\bar c_1|0}\equiv -i,$$ and it agrees with textbook conventions
\cite{polchinski}, to which we adhere in this paper.}
\be
\aver{0|c_{-1} c_0 c_1 \bar c_{-1}\bar c_0  \bar c_1|0}\equiv 1.
\ee
Computing the open string trace and using the usual modular
property of the Dedekind eta function
\be
\sqrt t\, \eta(it)=\eta\!\left(\frac i t\right),\label{modular}
\ee
 we find
\be
-4it\,\Tr_{H^{\rm ghost}_{\mathrm{open}}}\left[(-1)^F e^{-2\pi t
\left(L_0+\frac {26}{24}\right)} b_0 c_0\right]= 4t\,
\eta^2(it)=4\, \eta^2\!\left(\frac i t\right).
\ee
This gives
\be
{\cal N}_{gh}^2=1.
\ee
Notice how the scaling law (\ref{modular-scaling}) accounts for
the modular transformation (\ref{modular}).
We can fix the sign in ${\cal N}_{gh}$  by asking
\be
\label{bcBS-defexpr}
\aver{B_{bc}|(c_0-\bar c_0)|c\bar c}=\aver {(c_0-\bar c_0)c\bar
c(0)}_{\rm disk}.
\ee
The right hand side can be computed  by expressing
 $(c_0-\bar c_0)$ as a contour integral, mapping the disk to the
upper half plane and using the doubling trick. Normalizing the
basic ghost correlator in the usual way
\be
\aver{c(z_1)c(z_2)c(z_3)}=z_{12} z_{13} z_{23} \,,
\ee
where $z_{ij}\equiv z_i-z_j$, we find
\be
\aver {(c_0-\bar c_0)c\bar c(0)}_{\rm disk}=-2,
\ee
and the normalization of the ghost boundary state is thus given simply by
\be
{\cal N}_{gh}=1.
\ee


\newpage
\section{Some more lumps}
\label{a-lumps}
Here we collect the gauge invariant data of few
more lump solutions. All data have been obtained in the $(L,3L)$
scheme up to $L=12$.
\begin{itemize}
\item Single lump at $R=\sqrt3$
\end{itemize}
This is the same solution of MSZ \cite{MSZ} but in the $(L,3L)$
scheme.
\begin{figure}[h]
\begin{center}
\resizebox{3.5in}{1.6in}{\includegraphics{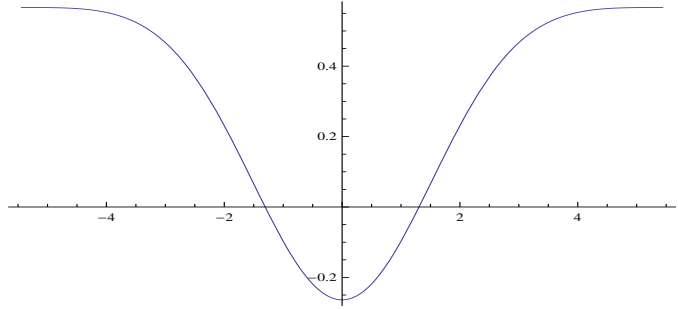}}
\end{center}
\caption{\label{fig:1lump-open} {\small Open string tachyon
profile of the MSZ single-lump solution obtained at $R=\sqrt3$ and
level $L=(12,36)$.}}
\end{figure}

{\footnotesize
\begin{center}
\begin{tabular}{|c|c|ccccccc|c|}
\hline
$L$           & ${\mathrm{Action}}$         & $E_0$      & $E_{1}$  &   $E_{2}$ &         $E_{3}$ &    $E_4$  &       $E_5$    &$E_6$            &$D$       \\
\hline
$ 1$  &                $ 1.32002$  &        $1.23951 $  & $0.74368 $  & $-$ &     $-       $ & $- $ &$-$ & $-$&                         $1.23951 $   \\
\hline
$ 2$  &                $ 1.09428$  &        $1.09094 $  & $0.830804 $  & $1.03277$ &     $-       $ & $- $ &$-$ & $-$&                         $-1.01353 $   \\
\hline
$ 3$  &                $ 1.06053$  &       $1.06017 $  & $0.905713  $  & $1.08758$  &     $1.36793$  & $-$ &$-$   &$-$                       &$-1.11078  $   \\
\hline
$4$      &            $1.03572$  &        $1.04623  $  & $0.918393  $  & $0.931471$  &    $1.4122$  &  $-     $   &    $-$    &$-$           &$-0.752479  $   \\
\hline
$5$  &               $1.02936$  &        $1.03948   $  & $0.94075  $  & $0.933722$ &      $0.678169$ &   $-       $  &    $-$  &$-$         &$-0.779229  $ \\
\hline
$6$        &         $1.02141$  &        $1.02921   $  & $0.946315  $  & $0.995166$ &    $0.676601$ & $2.06251$ &     $-$   &$-$            &$-0.945165  $ \\
\hline
$7$        &         $1.01868$  &        $1.02668   $  & $0.956761  $  & $0.996584$ &    $1.11184 $ & $2.11211$ &     $-$   &$-$            &$-0.959492 $ \\
\hline
$8$       &           $1.01454$  &        $1.02301  $  & $0.959784  $  & $0.977037$ &    $1.12839 $ & $-0.327725$ &   $-$  &$-$             &$-0.909528 $ \\
\hline
$9$       &           $1.01351$  &        $1.02171  $  & $0.965702  $  & $0.976881$ &    $0.859033$ & $-0.350675$ &   $3.66745 $  &$-$       &$-0.913363 $ \\
\hline
$10$       &           $1.01108$  &        $1.01787  $  & $0.967666 $  & $0.993958$ &    $0.860958 $ & $1.98806$&   $ 3.81063$  &$-$         &$-0.963774 $ \\
\hline
$11$       &           $1.01052$  &        $1.01708  $  & $0.971569 $  & $0.993933$ &    $1.04829 $ & $2.01551$&   $-4.09339$  &$-$           &$-0.966875  $ \\
\hline
$12$       &           $1.00893$  &        $1.01549  $  & $0.972933 $  & $0.98699$ &    $1.05428 $ & $-0.0353736$&   $-4.26896$ &$8.53484$    &$-0.945928  $ \\
\hline
Exp.       &           $1$  &               $1  $      & $1 $  & $1$ &    $1 $ & $1$&   $1$ &$1$    &$-1  $ \\
\hline
\end{tabular}
\end{center}         }

\newpage
\begin{itemize}
\item Single lump at $R=2\sqrt3$
\end{itemize}
This is a single lump centered at $x=\pi R$. Notice how this
reflects into alternating signs for the $E_n$ invariants.
\begin{figure}[h]
\begin{center}
\resizebox{3.5in}{1.6in}{\includegraphics{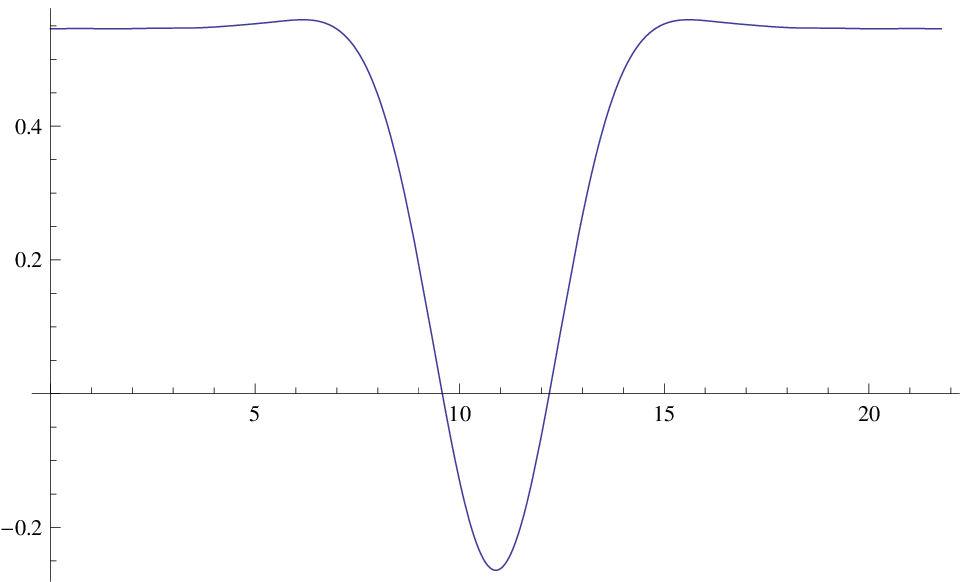}}
\end{center}
\caption{\label{fig:1lump-open} {\small Open string tachyon
profile of a single-lump solution obtained at $R=2\sqrt3$ and
level $L=(12,36)$.}}
\end{figure}

{\footnotesize
\begin{center}
\begin{tabular}{|c|c|c|cccccc|}
\hline
$L$& Action &  $D$  &   $E_0$    &  $E_1$     &    $E_2$ &   $E_3$ &    $E_4$  &   $E_5$ \\
\hline
1  &    1.84419&1.70138   &  1.70138     &  -- 0.72894           & 0.747337           & -- 0.723963          & --       &         --            \\
\hline
2  &    1.20098&-- 0.467964   &  1.30548     &  -- 0.77845           & 0.846539           & -- 0.91968          & 0.949864        &         --            \\
\hline
3  &    1.13151&-- 0.90771   &  1.25135     &  -- 0.880111          & 0.906433           & -- 0.928996         & 1.08763         &    -- 1.21329               \\
\hline
4  &    1.05813&-- 0.644043   &  1.1658      &  -- 0.893358          & 0.922018           & -- 0.945655         & 0.92909         &    -- 1.25847               \\
\hline
5  &    1.05079&-- 0.677223   &  1.15751     &  -- 0.926427          & 0.941257           & -- 0.963228         & 0.933319        &    -- 0.854891              \\
\hline
6  &    1.02895&-- 0.851623   &  1.11426     &  -- 0.931712          & 0.947518           & -- 0.970614         & 0.993906        &    -- 0.85989               \\
\hline
7  &    1.02724&-- 0.875745   &  1.11142     &  -- 0.945795          & 0.957258           & -- 0.970967         & 0.997313        &    -- 1.03384               \\
\hline
8  &    1.01773&-- 0.84127   &  1.09036     &  -- 0.948533          & 0.960557           & -- 0.974792         & 0.976872        &    -- 1.04075               \\
\hline
9  &    1.01724&-- 0.847968   &  1.08895     &  -- 0.957215          & 0.966076           & -- 0.979099         & 0.977358        &    -- 0.949906              \\
\hline
10 &    1.01217&-- 0.909139   &  1.07312     &  -- 0.958953          & 0.968161           & -- 0.981546         & 0.993873        &    -- 0.952527              \\
\hline
11 &    1.01204&-- 0.915063   &  1.07228     &  -- 0.964312          & 0.971903           & -- 0.981671         & 0.994434        &    -- 1.00923               \\
\hline
12 &    1.00897&-- 0.897553   &  1.06302     &  -- 0.965506          & 0.973333           & -- 0.983331         & 0.98709         &    -- 1.01181               \\
\hline
Exp.& 1 &-- 1 &      1     &     -- 1    &     1    &   -- 1   &     1     &    -- 1  \\
\hline
\end{tabular}
\end{center}
\begin{center}
\begin{tabular}{|c|ccccccc|}
\hline
$L$  &     $E_6$  &   $E_7$ & $E_8$     & $E_9$       & $E_{10}$  & $E_{11}$  & $E_{12}$ \\
\hline
1&     --          & --       & --         & --           & --         & --         & --         \\
\hline
2&     --          & --       & --         & --           & --         & --         & --         \\
\hline
3& 1.33609        & --       & --         & --           & --         & --         & --         \\
\hline
4& 1.38348        & --       & --         & --           & --         & --         & --         \\
\hline
5& 0.676515       &-- 1.67643 & --         & --           & --         & --         & --         \\
\hline
6& 0.677062       &-- 1.72458 &2.03206    & --           & --         & --         & --         \\
\hline
7& 1.1127         &-- 0.327417&2.12122    &-- 2.62668     & --         & --         & --         \\
\hline
8& 1.12425        &-- 0.320042&-- 0.314551  &-- 2.71207     & --         & --         & --         \\
\hline
9& 0.859104       &-- 1.3691  &-- 0.352809  &1.55456      &3.69837    & --         & --         \\
\hline
10&0.860785       &-- 1.3843  &1.97233    &1.61552      &3.78564    & --         & --         \\
\hline
11&1.04906        &-- 0.617564&2.01885    &-- 3.49068     &-- 4.12322   &-- 5.50346   & --          \\
\hline
12&1.05315        &-- 0.615319&-- 0.0261646 &-- 3.56055     &-- 4.23092   &-- 5.60153   & 8.48854         \\
\hline
Exp.&   1      &   -- 1      &    1       &      -- 1       &   1        &    -- 1       &      1           \\
\hline
\end{tabular}
\end{center}
}

\begin{itemize}
\item Symmetric double lump at $R=2\sqrt3$
This solution represents two D-branes at distance $a=\frac12$, as
defined in section~\ref{ss-R2sq3}. This double lump solution is
just the single lump solution we obtained at $R=\sqrt3$,
translated by half the period $\pi\sqrt3$ and periodically extended to circle of radius
$R=2\sqrt3$. This is clearly visible from the invariants, which, up
to the alternating signs, are exactly double of those for the single
lump at $R=\sqrt3$.

\begin{figure}[h]
\begin{center}
\resizebox{3.5in}{1.6in}{\includegraphics{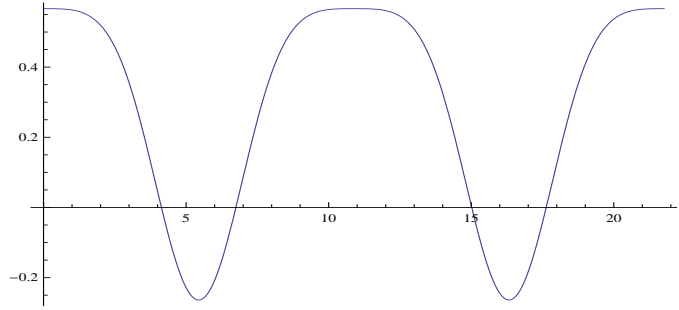}}
\end{center}
\caption{\label{fig:1lump-open} {\small Open string tachyon
profile of a symmetric double-lump solution obtained at
$R=2\sqrt3$ and level $L=(12,36)$.}}
\end{figure}

{\footnotesize
\begin{tabular}{|c|c|c|cccccc|}
  \hline
  $L$ & Action& $D$&     $E_0$ &$E_1$ &$E_2$ &$E_3$ &$E_4$    &$E_5$            \\
  \hline
 1 &2.64005&2.47902 & 2.47902& 0 & -- 1.48736 & --  & --   & --               \\
\hline
 2 &2.18856&-- 2.02706 & 2.18187& 0 & -- 1.66161 & 0 & 2.06554  & --               \\
\hline
 3 &2.12107&-- 2.22155 & 2.12035& 0 & -- 1.81143 & 0 & 2.17516  & 0               \\
\hline
 4 &2.07144&-- 1.50496 & 2.09245& 0 & -- 1.83679 & 0 & 1.86294  & 0               \\
\hline
 5 &2.05871&-- 1.55846 & 2.07896& 0 & -- 1.8815  & 0 & 1.86744  & 0               \\
\hline
 6 &2.04179&-- 1.89033 & 2.05843& 0 & -- 1.89263 & 0 & 1.99033  & 0               \\
\hline
 7 &2.03736&-- 1.91898 & 2.05336& 0 & -- 1.91352 & 0 & 1.99317  & 0               \\
\hline
 8 &2.02908&-- 1.81906 & 2.04602& 0 & -- 1.91957 & 0 & 1.95407  & 0               \\
\hline
 9 &2.02702&-- 1.82673 & 2.04341& 0 & -- 1.9314  & 0 & 1.95376  & 0               \\
\hline
 10&2.02216&-- 1.92755 & 2.03574& 0 & -- 1.93533 & 0 & 1.98792  & 0               \\
\hline
 11&2.02103&-- 1.93375 & 2.03417& 0 & -- 1.94314 & 0 & 1.98787  & 0               \\
\hline
 12&2.01785&-- 1.89186 & 2.03098& 0 & -- 1.94587 & 0 & 1.97398  & 0               \\
    \hline
 Expected  &2&    -- 2   &    2   &0 &   -- 2     & 0& 2        & 0                \\
  \hline
\end{tabular}

\begin{tabular}{|c|ccccccc|}
  \hline
  $L$    & $E_6$ & $E_7$ & $E_8$ & $E_9$ & $E_{10}$ & $E_{11}$ &$E_{12}$\\
\hline
  1      & --       & --  & --         & --  & --        & --  &-- \\
\hline
  2      & --       & --  & --         & --  & --        & --  &-- \\
  \hline
  3      & -- 2.73586 &--  & --          & --  & --        & --  &-- \\
  \hline
  4      & -- 2.82439 & --  & --          & --  & --        & --  &-- \\
  \hline
  5      & -- 1.35634 & 0 & --          & --  & --        & --  &-- \\
  \hline
  6      & -- 1.3532  & 0 & 4.12501   & --  & --        & --  &-- \\
  \hline
  7      & -- 2.22369 & 0 & 4.22422   & 0  & --        & --  &-- \\
  \hline
  8      & -- 2.25678 & 0 & -- 0.655451 & 0 & --        & --  &-- \\
  \hline
  9      & -- 1.71807 & 0 & -- 0.701349 & 0 & -- 7.3349 & --  &-- \\
  \hline
  10     & -- 1.72192 & 0 & 3.97612   & 0 & -- 7.62125& --  &-- \\
  \hline
  11     & -- 2.09657 & 0 & 4.03102   & 0 & 8.18679 & 0 &-- \\
  \hline
  12     & -- 2.10857 & 0 &-- 0.0707473 & 0 & 8.53792 & 0 &17.0697\\
  \hline
  Expected &    -- 2    & 0 & 2         & 0 & -- 2      & 0 &2\\
  \hline
\end{tabular}
}
\end{itemize}

\end{appendix}


\begin{thebibliography}{99}
\bibitem{Witten}
  E.~Witten,
  ``Noncommutative Geometry And String Field Theory,''
  Nucl.\ Phys.\  B {\bf 268}, 253 (1986).


\bibitem{TZ}
  W.~Taylor and B.~Zwiebach,
  ``D-branes, tachyons, and string field theory,''
  arXiv:hep-th/0311017.

\bibitem{FK}
  E.~Fuchs and M.~Kroyter,
  ``Analytical Solutions of Open String Field Theory,''
  arXiv:0807.4722 [hep-th].

\bibitem{Sen:2004cq}
  A.~Sen,
  ``Energy momentum tensor and marginal deformations in open string field theory,''
   JHEP {\bf 0408} (2004) 034  [hep-th/0403200].  

\bibitem{Ellwood}
  I.~Ellwood,
  ``The Closed string tadpole in open string field theory,''
  JHEP {\bf 0808} (2008) 063
  [arXiv:0804.1131 [hep-th]].

\bibitem{KOZ}
  M.~Kiermaier, Y.~Okawa and B.~Zwiebach,
  ``The boundary state from open string fields,''
  arXiv:0810.1737 [hep-th].



\bibitem{KKT}
  T.~Kawano, I.~Kishimoto and T.~Takahashi,
  ``Gauge Invariant Overlaps for Classical Solutions in Open String Field Theory,''
  Nucl.\ Phys.\ B\ {\bf 803} (2008) 135
  [arXiv:0804.1541 [hep-th]].

\bibitem{Ishibashi}
  N.~Ishibashi,
  ``The Boundary and Crosscap States in Conformal Field Theories,''  Mod.\ Phys.\ Lett.\ A {\bf 4} (1989) 251.  


\bibitem{Sen:1999xm}
  A.~Sen,
  ``Universality of the tachyon potential,''  JHEP {\bf 9912} (1999) 027  [hep-th/9911116].  


\bibitem{cons}
  L.~Rastelli and B.~Zwiebach,
  ``Tachyon potentials, star products and universality,''  JHEP {\bf 0109} (2001) 038  [hep-th/0006240].  

\bibitem{Takahashi:2011wk}
  D.~Takahashi,
  ``The boundary state for a class of analytic solutions in open string field theory,''  JHEP {\bf 1111} (2011) 054  [arXiv:1110.1443 [hep-th]].  



\bibitem{Rozali}
  A.~Rajaraman and M.~Rozali,
  ``D-branes in linear dilaton backgrounds,''  JHEP {\bf 9912} (1999) 005  [hep-th/9909017].  

\bibitem{Gaberdiel}
M.~Gaberdiel, ``Boundary conformal field theory and D-branes'',
Lectures given at the TMR network school on �Nonperturbative
methods in low dimensional integrable models�, Budapest, 15-21
July 2003. http://www.phys.ethz.ch/~mrg/lectures2.pdf




\bibitem{KO}
  M.~Kiermaier and Y.~Okawa,
  ``Exact marginality in open string field theory: A General framework,''  JHEP {\bf 0911} (2009) 041  [arXiv:0707.4472 [hep-th]].  

\bibitem{Fuchs:2007yy}
  E.~Fuchs, M.~Kroyter and R.~Potting,
  ``Marginal deformations in string field theory,''  JHEP {\bf 0709} (2007) 101  [arXiv:0704.2222 [hep-th]].  


\bibitem{BMT}
  L.~Bonora, C.~Maccaferri and D.~D.~Tolla,
  ``Relevant Deformations in Open String Field Theory: a Simple Solution for Lumps,''  JHEP {\bf 1111} (2011) 107  [arXiv:1009.4158 [hep-th]].  

\bibitem{multibranes}
  M.~Murata and M.~Schnabl,
  ``Multibrane Solutions in Open String Field Theory,'' JHEP {\bf 1207} (2012) 063  [arXiv:1112.0591 [hep-th]].  

\bibitem{marg1}
  M.~Schnabl,
  ``Comments on marginal deformations in open string field theory,''
  Phys.\ Lett.\ B\ {\bf 654} (2007) 194
  [hep-th/0701248 [HEP-TH]].

\bibitem{marg2}
  M.~Kiermaier, Y.~Okawa, L.~Rastelli and B.~Zwiebach,
  ``Analytic solutions for marginal deformations in open string field theory,''
  JHEP\ {\bf 0801} (2008) 028
  [hep-th/0701249 [HEP-TH]].

\bibitem{Erler:2007rh}
  T.~Erler,
  ``Marginal Solutions for the Superstring,''  JHEP {\bf 0707} (2007) 050  [arXiv:0704.0930 [hep-th]].  

\bibitem{Kiermaier:2010cf}
  M.~Kiermaier, Y.~Okawa and P.~Soler,
  ``Solutions from boundary condition changing operators in open string field theory,''  JHEP {\bf 1103} (2011) 122  [arXiv:1009.6185 [hep-th]].  

\bibitem{Schnabl}
  M.~Schnabl,
  ``Analytic solution for tachyon condensation in open string field theory,''
  Adv.\ Theor.\ Math.\ Phys.\  {\bf 10} (2006) 433
  [arXiv:hep-th/0511286].



\bibitem{Okawa}
  Y.~Okawa,
  ``Comments on Schnabl's analytic solution for tachyon condensation in
  Witten's open string field theory,''
  JHEP {\bf 0604} (2006) 055
  [arXiv:hep-th/0603159].

\bibitem{Erler1}
  T.~Erler,
  ``Split string formalism and the closed string vacuum,''
  JHEP {\bf 0705} (2007) 083
  [arXiv:hep-th/0611200].

\bibitem{Erler2}
  T.~Erler,
  ``Split string formalism and the closed string vacuum. II,''
  JHEP {\bf 0705} (2007) 084
  [arXiv:hep-th/0612050].

\bibitem{lightning}
  M.~Schnabl,
  ``Algebraic solutions in Open String Field Theory - a lightning review,''
  arXiv:1004.4858 [hep-th].


\bibitem{Kishimoto:2008zj}
  I.~Kishimoto,
  ``Comments on gauge invariant overlaps for marginal solutions in open string field theory,''
  Prog.\ Theor.\ Phys.\  {\bf 120} (2008) 875  [arXiv:0808.0355 [hep-th]].  


\bibitem{Noumi:2011kn}
  T.~Noumi and Y.~Okawa,
  ``Solutions from boundary condition changing operators in open superstring field theory,''
  JHEP {\bf 1112} (2011) 034  [arXiv:1108.5317 [hep-th]].  


\bibitem{Erler:2012qr}
  T.~Erler and C.~Maccaferri,
  ``The Phantom Term in Open String Field Theory,''  JHEP {\bf 1206} (2012) 084
  [arXiv:1201.5122 [hep-th]]. 


\bibitem{Erler:2012qn}
  T.~Erler and C.~Maccaferri,
  ``Connecting Solutions in Open String Field Theory with Singular Gauge Transformations,''
   JHEP {\bf 1204} (2012) 107  [arXiv:1201.5119 [hep-th]].  

\bibitem{sen-review}
  A.~Sen,
  ``Tachyon dynamics in open string theory,''  Int.\ J.\ Mod.\ Phys.\ A {\bf 20} (2005) 5513  [hep-th/0410103].  


\bibitem{Larsen}
  F.~Larsen, A.~Naqvi and S.~Terashima,
  ``Rolling tachyons and decaying branes,''  JHEP {\bf 0302} (2003) 039  [hep-th/0212248].  

\bibitem{ell-roll}
  I.~Ellwood,
  ``Rolling to the tachyon vacuum in string field theory,''  JHEP {\bf 0712} (2007) 028  [arXiv:0705.0013 [hep-th]].  

\bibitem{Hellerman:2008wp}
  S.~Hellerman and M.~Schnabl,
  ``Light-like tachyon condensation in Open String Field Theory,''  arXiv:0803.1184 [hep-th].  


\bibitem{MSZ}
  N.~Moeller, A.~Sen, B.~Zwiebach,
  ``D-branes as tachyon lumps in string field theory,''
  JHEP {\bf 0008 } (2000)  039.
  [hep-th/0005036].


\bibitem{Moeller}
  N.~Moeller,
  ``Codimension two lump solutions in string field theory and tachyonic theories,''
  [hep-th/0008101].


\bibitem{Beccaria}
  M.~Beccaria,
  ``D0-brane tension in string field theory,''  JHEP {\bf 0509} (2005) 021  [hep-th/0508090].  

\bibitem{thesis}
 A.~Kurs,
 ``Classical solutions in string field theory,''
Senior thesis, Princeton University, 2005


\bibitem{KS}
M.~Kudrna, M.~Schnabl, {\it to appear}

\bibitem{japan}
  M.~Kudrna, T.~Masuda, Y.~Okawa, M.~Schnabl and K.~Yoshida,
  ``Gauge-invariant observables and marginal deformations in open string field theory,''
  arXiv:1207.3335 [hep-th].


\bibitem{Erler:2010pr}
  T.~Erler,
  ``Exotic Universal Solutions in Cubic Superstring Field Theory,''  JHEP {\bf 1104} (2011) 107  [arXiv:1009.1865 [hep-th]].  



\bibitem{Large}
  A.~Sen and B.~Zwiebach,
  ``Large marginal deformations in string field theory,''  JHEP {\bf 0010} (2000) 009  [hep-th/0007153].  



\bibitem{Canadians}
  J.~L.~Karczmarek and M.~Longton,
  ``SFT on separated D-branes and D-brane translation,''  arXiv:1203.3805 [hep-th].  

\bibitem{Kausch:1995py}
  H.~G.~Kausch,
  ``Curiosities at c = -2,''
  hep-th/9510149.

\bibitem{Gaiotto:2003yb}
  D.~Gaiotto and L.~Rastelli,
  ``A Paradigm of open / closed duality: Liouville D-branes and the
Kontsevich model,''
  JHEP {\bf 0507} (2005) 053
  [hep-th/0312196].

\bibitem{wedges}
  M.~Schnabl,
  ``Wedge states in string field theory,''  JHEP {\bf 0301} (2003) 004  [hep-th/0201095].  

\bibitem{Leclair}
  A.~LeClair, M.~E.~Peskin and C.~R.~Preitschopf,
  ``String Field Theory on the Conformal Plane. 1. Kinematical Principles,''  Nucl.\ Phys.\ B {\bf 317} (1989) 411.  

\bibitem{barton}
  B.~Zwiebach,
  ``Closed string field theory: Quantum action and the B-V master equation,''  Nucl.\ Phys.\ B {\bf 390} (1993) 33  [hep-th/9206084].  

\bibitem{polchinski}
  J.~Polchinski,
  ``String theory. Vol. 1: An introduction to the bosonic string,''  Cambridge, UK: Univ. Pr. (1998) 402 p

\end{thebibliography}
\end{document}